\documentclass[aps,prb,twocolumn,showpacs]{revtex4-1}
\usepackage{bm}
\usepackage{color} 
\usepackage{graphicx}
\usepackage{epsfig}
\usepackage{hyperref}
\usepackage{natbib}
\usepackage{amsmath}
\usepackage{amssymb}

\newcommand{\rmi}{\mathrm{i}}
\newcommand{\e}{\mathrm{e}}
\newcommand{\eiwt}{\e^{-\rmi \omega_0 t}}
\newcommand{\tf}{t_{\rm f}}

\newcommand{\f}{f_{hh'}}

\begin{document}

\title{Reflection of short polarized optical pulses from periodic and aperiodic\\ multiple quantum well structures}

\author{A.\,V.\,Poshakinskiy}\email{poshakinskiy@mail.ioffe.ru}
\author{A.\,N.\,Poddubny}
\author{S.\,A.\,Tarasenko}
\affiliation{Ioffe Physical-Technical Institute, Russian Academy of Sciences, 194021 St.~Petersburg, Russia}
\begin{abstract}
We study the reflection of polarized optical pulses from resonant photonic structures
formed by periodic, Fibonacci, and gradient sequences of quantum wells. The form and polarization of the reflected pulse are shown to be determined by the structure design and optical length. In structures with periodic quantum well arrangement, the response to ultrashort pulse is an optical signal with a sharp rise followed by an exponential decay or Bessel beats
depending on the structure length. The duration of reflected pulses non-monotonically depends on the number of quantum wells  reaching the minimum for a certain structure length which corresponds to the transition from superradiant to photonic-crystalline regime. We also study the conversion of pulse polarization in the longitudinal external magnetic field which splits the exciton resonance. Comparing periodic, Fibonacci, and gradient structures we show that the latter are more efficient for the conversion from linear to circular polarization.
\end{abstract}

\maketitle

\section{Introduction}

The optical spectroscopy of photonic structures, where the susceptibility is spatially modulated on the scale of light wavelength, has been attracting a lot of attention due to fundamental interest in the light-matter coupling as well as applications of such systems to optical devices.\cite{Limonov_book,Ivchenko_book}
The photonic structures sharing similar physics can be realized in different systems ranging from crystals for $\gamma$-rays,\cite{kagan1999,Shvydko2010} solid state heterostructures for infrared,  visible, or ultraviolet light\cite{Feltin2006,Megens2009} to the
optical lattices of laser-cooled atoms~\cite{deutsch1995} and the chains of optical cavities.\cite{yanik2004} 
Of special interest in this field are the resonant photonic structures based on periodically arranged 
semiconductor quantum wells (QWs) or dots (QDs) where the light couples to excitons confined in the QWs or QDs
leading to formation of polaritonic modes.\cite{Kavbamalas_book} An advantage of such structures is the possibility of electric control of the light-exciton coupling as well as tuning the spectral position of the exciton resonance.\cite{prineas2006apl,chaldyshev2011}

The QW-based resonant photonic crystals suggested two decades ago\cite{Ivchenko1994} are extensively studied at present both  theoretically~\cite{koch1996mqw,Cho2002,Pilozzi2004,Voronov2004,Averkiev2012} and experimentally~\cite{Kochereshko1994,Merle1996,Prineas2000,Goldberg2009,chaldyshev2011,iorsh2011,chaldyshev2011b}
with the focus on their spectral characteristics. The progress in the optical spectroscopy technique makes it possible to study the response of photonic structures at the pico- and femtosecond time scale,\cite{ammerlahn2000,Volz2012}  which stimulates theoretical investigations of ultrafast optical processes.
In the present paper, we develop the theory of time- and polarization- resolved linear response of 
multiple QW systems.
We show that the optical spectroscopy in the time domain, being complementary to that in the frequency domain, provides direct information on the lifetime and coherence of optical excitations,
which cannot be easily obtained from the stationary reflection spectra. Previous studies of time-dependent reflectivity were limited to periodic QW structures.\cite{ammerlahn2000,Andreani1998,koch1996mqw} Here, we focus on the effect of geometrical arrangement of QWs on the optical response and compare the responses of periodic, quasiperiodic Fibonacci\cite{Poddubny2008prb,Hendrickson2008}  as well as gradient QW structures.
We also analyze the linear-to-linear and linear-to-circular polarization conversion of pulses which occurs in the external magnetic field.

The paper is organized as follows. The multiple QW structures and the general expressions for the structure optical response are presented in Sec.~\ref{sec:model}. In Secs.~\ref{sec:periodic} and~\ref{sec:abs}, we study the reflection of short optical pulses from periodic and aperiodic Bragg structures, respectively. The polarization conversion is analyzed in Sec.~\ref{sec:polarized}. 
The results of the paper are summarized in Sec.~\ref{sec:summary}.

\section{Model} \label{sec:model}
\begin{figure}[b]
\includegraphics[width=0.45\textwidth]{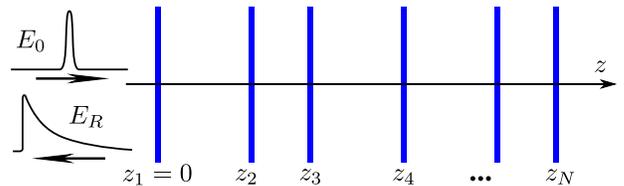}
\caption{(Color online) Illustration of optical pulse reflection from a multiple quantum well structure.}
\label{figure1}
\end{figure}

We consider the layered heterostructure which consists of $N$ identical quantum wells embedded in a dielectric matrix, see Fig.~\ref{figure1}. The wells are centered at the points $z_j$ ($j=1 \ldots N$) with $z_1=0$. The structure is characterized by the background refractive index $n_b$. Additionally, the linear response of each QW to the electromagnetic field has the exciton resonance at the frequency $\omega_0$. The corresponding frequency dependence of the amplitude reflection and transmission coefficients, $r_{1}(\omega)$ and $t_{1}(\omega)$, respectively, is given by
\begin{equation}
r_{1} (\omega) = - \frac{\rmi \Gamma_0}{\omega - \omega_0 + \rmi (\Gamma_0+\Gamma)} \:, \; t_{1} (\omega) = 1 + r_{1} (\omega) \:,
\end{equation}
where $\Gamma_0$ and $\Gamma$ are the radiative and non-radiative decay rates. We assume that the QWs are all identical and uniform in the interface plane, so that the effects of inhomogeneity\cite{Andreani1998} can be neglected.

The light pulse is normally incident on the QW structure along the $z$ axis and reflected. In the left half-space, $z<0$, the electric field $\bm{E}(z,t)$ is the sum of incident and reflected waves,
\begin{equation}\label{E_total}
\bm{E}(z,t) = \bm{E}_0(t-zn_b/c) + \bm{E}_R(t+zn_b/c)  \:,
\end{equation}
where $c$ is the speed of light. The reflected pulse $\bm{E}_R(t)$ is determined by 
the linear response function $\rho(\tau)$ as follows: 
\begin{equation}\label{E_R_time}
\bm{E}_R(t)=\int_{0}^{\infty} \rho(\tau) \bm{E}_{0}(t-\tau) d\tau \:.
\end{equation}
The function $\rho(\tau)$, in turn, represents the Fourier transformation of the reflection coefficient of the structure in the frequency domain $r(\omega)$,
\begin{equation}\label{eq:rho}
\rho(t) = \int_{-\infty}^{\infty} r(\omega) {\rm e}^{-\rmi \omega t} d\omega /(2\pi) \:.
\end{equation}

We calculate the reflected pulse $\bm{E}_R(t)$ after Eqs.~(\ref{E_R_time}) and~(\ref{eq:rho}). The reflection coefficient for an arbitrary QW structure is found by the transfer-matrix technique.\cite{Ivchenko_book} In this approach, each layer of the structure is described by the transfer matrix which connects the amplitudes of incoming and outgoing waves. The corresponding matrices for an individual QW and interwell distance $d$ are given by 
\begin{equation}
T_{QW} = \frac{1}{t_{1}(\omega)}
\begin{bmatrix}
t_{1}^2(\omega) - r_{1}^2(\omega) \; & r_{1}(\omega) \\ - r_{1}(\omega) & 1
\end{bmatrix}
\:,
\end{equation}
\begin{equation}
T_{d} = 
\begin{bmatrix}
\exp(\rmi \omega d n_b /c) & 0 \\ 0 & \exp(-\rmi \omega d n_b /c)
\end{bmatrix}
\:,
\end{equation}
respectively. The total transfer matrix of the structure $T$ is given by the product of the individual transfer matrices of all layers. It enables one to calculate the reflection $r(\omega)$ and transmission $t(\omega)$ coefficients from the matrix equation
\begin{equation}
T 
\begin{bmatrix}
1 \\ r(\omega)
\end{bmatrix}
=
\begin{bmatrix}
t(\omega) \\ 0
\end{bmatrix} 
\:.
\end{equation}
 
In what follows we are interested in the optical response to ultrashort pulses which have the carrier frequency at the resonance frequency $\omega_0$ and the spectral width $\delta\omega$ much broader than the scale of the exciton-related spectral features, $\sqrt{\omega_0\Gamma_0} \ll \delta\omega \ll \omega_0$. Since the typical band width of exciton reflection in 
periodic QW structures is of the order of few meV,\cite{ammerlahn2000,Hendrickson2008,chaldyshev2011} this condition can be realized for sub-picosecond pulses. Then, the time dependence of the reflected pulse electric field has the form $\bm{E}_R(t) = \rho(t) \int_{-\infty}^{+ \infty} \bm{E}_0 (\tau) {\rm e}^{{\rm i}\omega_0 \tau} d\tau$ and follows $\rho(t)$.  

\section{Periodic Bragg structures} \label{sec:periodic}

In this Section, we consider periodic QW structures where the interwell distance $d$ equals to half of the light wavelength at the exciton resonance frequency
\begin{equation} \label{eq:bragg}
d=\pi c /(\omega_0 n_b) \:.
\end{equation} 
The calculation by the transfer matrix technique shows that the reflection coefficient of such structure containing $N$ QWs in the frequency range $|\omega-\omega_0|\ll \omega_0$ can be compactly presented by\cite{Ivchenko1994,Ivchenko_book}
\begin{eqnarray}\label{eq:pnw}
r_N(\omega)=\left( 2w\tilde w-1+2 \rmi\sqrt{w\tilde w}\sqrt{w\tilde w-1} \cot \Phi \right)^{-1} \:.
\end{eqnarray}
Here, $w=(\omega-\omega_0)/\Delta$, $\tilde w =  w + \rmi \Gamma/\Delta$, $\Delta=\sqrt{(2/\pi)\omega_0\Gamma_0}$ is the halfwidth of the stop band,
$\Phi=2\nu\sqrt{w^2-w/\tilde w}$ is a resonant contribution to light phase incursion,
and $\nu=N\Gamma_0/\Delta$ is the normalized number of QWs.

\begin{figure}[ht]
\includegraphics[width=0.49\textwidth]{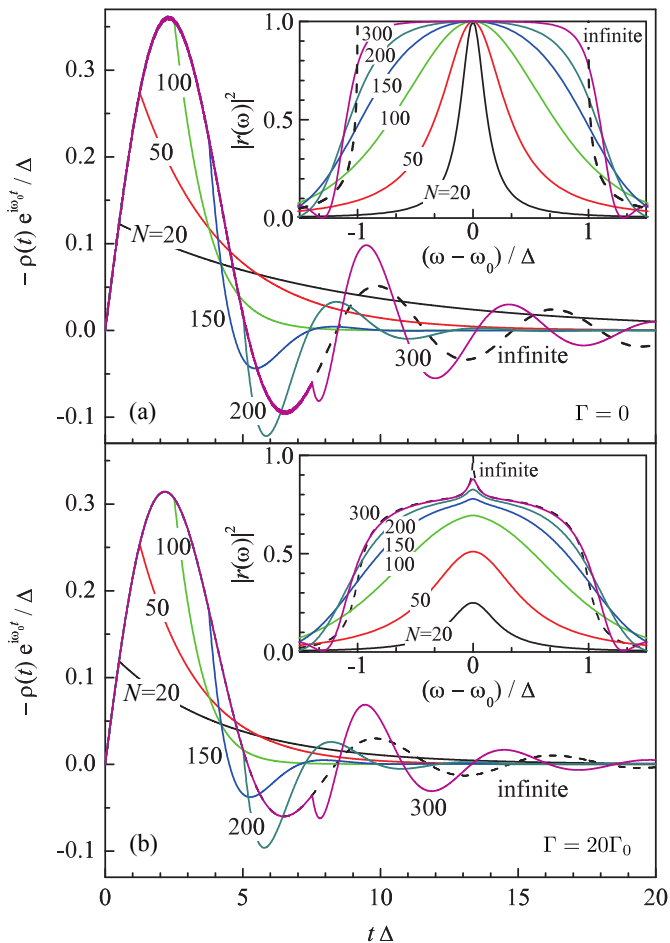}
\caption{(Color online) The envelopes of reflected pulses calculated for the resonant Bragg structures with various numbers of QWs and $\omega_0/\Gamma_0 = 4 \times 10^4$. Figures~(a) and~(b) correspond to the absence, $\Gamma=0$, and rather strong, $\Gamma=20\Gamma_0$, exciton non-radiative decay, respectively. The dashed curves depict the reflected pulse envelopes for the semi-infinite Bragg structure. Insets show the reflection spectra of the same QW structures in the frequency domain.}
\label{figure2}
\end{figure}

Figure~\ref{figure2} shows the envelopes of reflected pulses in the time domain calculated numerically for structures with the various QW number. The curves in Figs.~\ref{figure2}(a) and~\ref{figure2}(b) are plotted, respectively, for the cases of zero and rather strong non-radiative decay. One can see that the shapes of reflected signals in both cases are similar and drastically depend on the QW structure length. In short structures, the pulse envelope has a sharp rise followed by a slow monotonous decay. Both the signal amplitude and the decay rate increase with the QW number $N$. In contrast, the reflected pulse envelope in long and, particular, semi-infinite structures exhibits oscillations. The period of the oscillations is determined by the stop band width $\Delta$. The calculation shows that, in the absence of non-radiative decay, the  transition between these regimes occurs at $\nu \approx 0.66$, which corresponds to $N=106$ for QW parameters given in the caption of Fig.~\ref{figure2}(a).
For comparison, we plot the reflection spectra $|r(\omega)|^2$ of the QW structures in the insets of Figs.~\ref{figure2}(a) and~\ref{figure2}(b). With the increase in QW number, the reflection spectrum gradually evolves  from the Lorentzian to a silk-hat profile without prominent changes. It demonstrates that the time-resolved spectroscopy can be an efficient tool to study photonic structures supplementing the frequency-resolved approach. 
In the rest of Section, we present analytical results for short and long periodic structures and study the transition between these two regimes of light reflection. We also analyze the role of non-radiative decay. 

\subsection{Transition from superradiant to photonic-crystalline regime}\label{ssec:tspcr}

The optical response of resonant Bragg structures with the small number of layers, $\nu \ll 1$ or $N \ll \sqrt{\omega_0 / \Gamma_0} \sim 100$ for typical QWs, is described by two times: the short flight time of light through the structure and back $t_{\rm f}=2\pi N/\omega_0$  
and the long lifetime of QW excitation. During its flight through the structure, the incident pulse sequentially excites QWs. Therefore, the reflected signal amplitude linearly increases with time at $t<t_{\rm f}$, see Figs.~\ref{figure2}(a) and~\ref{figure2}(b). The signal behavior at the times $t \gg t_f$ is determined by the excitation decay. To describe it analytically one should consider the limit $|w|, |\tilde{w}| \ll 1$ in Eq.~(\ref{eq:pnw}), which yields the Lorentzian for the reflection coefficient in the frequency domain,\cite{Ivchenko1994}
\begin{equation}\label{eq:rsr}
r_{N}(\omega)=
-\frac{\rmi N\Gamma_0}{\omega-\omega_0+\rmi(N\Gamma_0+\Gamma)} \:.
\end{equation}
The corresponding response function $\rho_{N}(t)$ determining the time dependence of reflected signal has the form
\begin{equation}\label{eq:rhosr}
\rho_{N}(t)=-N\Gamma_0\e^{-(N\Gamma_0+\Gamma)t}  \eiwt \, \theta(t) \:,
\end{equation} 
where $\theta(t)$ is the Heaviside step function. The pulse envelope exponentially decays with time at the rate proportional to the QW number.\cite{koch1996mqw,Andreani1998} Such an enhancement of the decay rate is a direct manifestation of the Dicke superradiance.\cite{Dicke1954,Khitrova2007nat}

The further growth of the structure length leads to the increase of the flight time and decrease of the decay time. At $N \sim \sqrt{\omega_0 /\Gamma_0}$, both times become comparable to each other and the optical response qualitatively changes. This corresponds to the transition from superradiant to photonic-crystalline regime with the stop band formation. In the limiting case of $N \gg \sqrt{\omega_0 /\Gamma_0}$, the structure reflectivity $|r(\omega)|^2$ is close to unity in the stop band spectral range $|\omega-\omega_0|<\Delta$ and rapidly decreases outside this stop band, see inset in Fig.~\ref{figure2}(a). The reflection coefficient of the semi-infinite structure is given by 
\begin{eqnarray}\label{eq:rinf}
r_{\infty}(\omega)&=&\frac{1}{\left( \sqrt{w\tilde w}+\sqrt{w\tilde w-1} \right)^2} \:.
\end{eqnarray}
The corresponding response function in the time domain at $\Gamma \ll \Delta$ and $t \gg \Gamma^2 / \Delta^3$ has the form
\begin{equation}\label{eq:rhoinfgamma}
\rho_{\infty}(t) = - \frac{2}{t} \left[ J_2(\Delta t) + \frac{\Gamma}{2\Delta} I_1(\Gamma t/2) \right] \e^{-\Gamma t/2} \eiwt \theta(t) \:,
\end{equation}
where $J_2$ and $I_1$ are the Bessel and modified Bessel functions of the first kind. In particular, in the absence of non-radiative decay, $\rho_{\infty}(t)$ is simplified to 
\begin{equation}\label{eq:rhoinf}
\rho_{\infty}(t) = -\frac{2}{t} J_2(t \Delta) \eiwt \, \theta(t) \:.
\end{equation}
Equations~(\ref{eq:rhoinfgamma}) and~(\ref{eq:rhoinf}) demonstrate that the envelope of the pulse reflected from a long structure oscillates in time. The oscillation frequency is determined by the stop band halfwidth $\Delta$. 
We note that oscillations of such kind are known for the 
scattering of $\gamma$-rays by nuclei.\cite{kagan1999} However, to the best of our knowledge, they have not been considered so far for multiple QW structures. 

\begin{figure}[t!]
\includegraphics[width=0.45\textwidth]{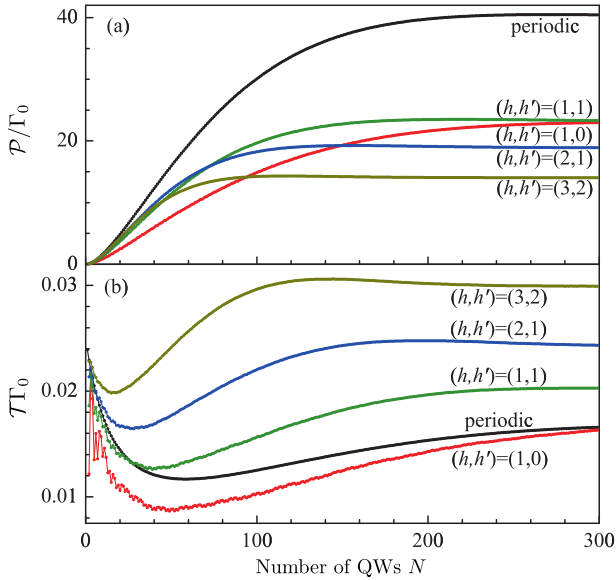}
\caption{(Color online) Dependences of the normalized energy $\mathcal{P}_N$ [figure~(a)] and duration $\mathcal{T}_N$ [figure~(b)] of the reflected pulse on the number of QWs. The curves are plotted for the periodic Bragg structures and the Fibonacci structures with different Bragg conditions, $\omega_0/\Gamma_0 = 4 \times 10^4$, and $\Gamma=20\Gamma_0$.}
\label{figure3}
\end{figure}

The transition from superradiant to photonic-crystalline regime can be also revealed by studying the
normalized energy $\mathcal{P}_N$ and duration $\mathcal{T}_N$ of the reflected pulse which we define by
\begin{equation}\label{eq:PT}
\mathcal{P}_N=\int_0^{\infty} |\rho(t)|^2 dt \:,\quad 
\mathcal{T}_N=\frac{1}{\mathcal{P}_N} \int_0^{\infty} |\rho(t)|^2 \, t \, dt \:.
\end{equation}
The dependences of $\mathcal P_N$ and $\mathcal T_N$ on the number of QWs $N$ for periodic structures are shown in Figs.~\ref{figure3}(a) and~\ref{figure3}(b), respectively, by black curves. 
The reflected energy increases with the QW number in the superradiant regime, $\mathcal P_N = N^2 \Gamma_0^2 /[2(N \Gamma_0 + \Gamma)]$, and reaches the value of $\mathcal P_{\infty} = 16 \Delta/(15\pi)$ in the photonic-crystalline regime provided $\Gamma \ll \Delta$. The transition from the linear growth to saturation of $\mathcal P_N$ corresponds to the transition from superradiant to photonic-crystalline regime. The dependence of the reflected pulse duration $\mathcal T_N$ on the QW number $N$ is even more pronounced. In the superradiant regime, it decreases with QW number as $\mathcal{T}_N = 1/(2N\Gamma_0 + 2\Gamma)$. In the area of the superradiant-photonic-crystalline regime transition $\mathcal T_N$ reaches the minimum $\approx 2.1 /\Delta$ at $N \approx  0.51 \Delta/\Gamma_0$ (for $\Gamma \ll \Delta$) and then again increases with the structure length. Finally, it saturates at the value $\mathcal{T}_{\infty} = 15 \pi / (16 \Delta)$ (for $\Gamma \ll \Delta$) for the semi-infinite structure. Numerical calculation shows that, for the QW parameters given in the caption to Fig.~\ref{figure3}, the shortest time $\mathcal T_N$ is achieved in the structure with $N = 58$. This provides a receipt for designing the fast-reflecting Bragg structures: the structure length should be intermediate and correspond to the transition between superradiant and photonic-crystalline regimes.

\subsection{Long structures. Echo and Borrmann effects.}\label{ssec:borrmann}

Now we discuss the features of pulse reflection from long, $N \gg \sqrt{\omega_0 /\Gamma_0}$, but finite QW structures.
The reflection coefficient of such structures $r_N(\omega)$ is described by Eq.~\eqref{eq:rinf} with corrections which can be obtained by expanding Eq.~(\ref{eq:pnw}) in the geometrical series in the parameter $\exp(2\rmi \Phi)$. Such a procedure yields
\begin{equation}\label{eq:reflsum}
r_N(\omega) = r_{\infty}(\omega) + \sum_{p=1}^{\infty} \delta r_N^{(p)}(\omega) \:,
\end{equation}
where
\begin{eqnarray}
\delta r_N^{(p)}(\omega)&=& -\frac{4\sqrt{w\tilde w}\sqrt{w\tilde w-1}}{\left[ \sqrt{w\tilde w}+\sqrt{w\tilde w-1} \right]^{4p}} \exp(2\rmi \Phi p) \:.
\end{eqnarray}
The corresponding response function in the time domain $\rho_N(t)$ has the form
\begin{equation}
\rho_N(t)=\rho_{\infty}(t)+\sum_{p=1}^{\infty} \delta \rho_N^{(p)}(t) \:,
\end{equation}
with $\delta \rho_N^{(p)}(t)$ being the Fourier image of $\delta r_N^{(p)}(\omega)$. The correction $\delta \rho_N^{(p)}(t)$ can be interpreted as the contribution to the response function caused by $p$-th reflection of the optical pulse from the structure back edge. 
Indeed, $\delta \rho_N^{(p)}(t) \neq 0$ only for $t \geq pt_{\rm f}$, where $pt_{\rm f}=2\pi N p/\omega_0$ is the time required for the light to travel through the structure and back $p$ times. In particular, in the absence of non-radiative decay, the corrections $\delta \rho_N^{(p)}(t)$ have the approximate form 
\begin{multline}
\rho_N^{(p)}(t) \approx - 8 \frac{(t- p \tf)^{2p-7/4}  (p \tf)^2 \, t}{(t+p\tf)^{2p + 7/4}} \sqrt{\frac{\Delta}{2\pi}}
\\\times\cos \left[ \frac{\pi}{4} + \Delta \sqrt{t^2 - (p \tf)^2} \right] \eiwt \theta(t-p\tf)  \:,
\end{multline}
which is obtained by the stationary phase method. 

\begin{figure}[ht]
\includegraphics[width=0.45\textwidth]{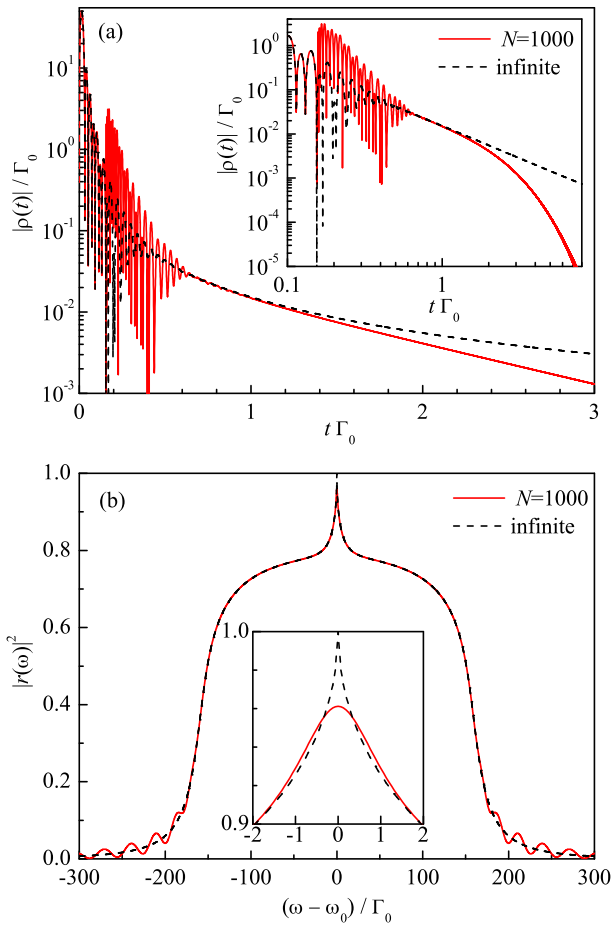}
\caption{(Color online) (a) The envelopes of the pulses reflected from the periodic Bragg structure with $1000$ QWs (red solid curve) and semi-infinite structure (black dashed curve). The curves are calculated for $\omega_0/\Gamma_0 = 4 \times 10^4$ and $\Gamma=20\Gamma_0$. The inset shows the envelopes plotted in the double logarithmic scale. (b) Reflection spectra of the same QW structures in the frequency domain. The inset shows the reflection spectrum details at $\omega \approx \omega_0$.}
\label{figure4}
\end{figure}

The first-order reflection from the structure back edge is clearly seen in the pulse envelopes plotted for the QW structures with $N=300$ in Figs.~\ref{figure2}(a) and~\ref{figure2}(b). It appears at the time $t=\tf$ [$t\Delta \approx 7.5$ in Figs.~\ref{figure2}(a) and~\ref{figure2}(b)] 
leading to an
uprise of the pulse envelope. This echo-like feature is more pronounced in Fig.~\ref{figure4}(a) where we compare the response functions $\rho_N(t)$ for long ($N=10^3$ or $\nu \approx 6.3$, red solid curve) and semi-infinite (black dashed curve) QW structures. The curves in Fig.~\ref{figure4}(a) are all calculated for structures with the strong non-radiative decay.

Another interesting difference between the response functions of a long finite and the semi-infinite QW structures is their behavior at large times, compare solid and dashed curves in Fig.~\ref{figure4}(a). 
The response function of the semi-infinite structure in the presence of non-radiative decay is described by Eq.~(\ref{eq:rhoinfgamma}). The counterintuitive feature of Eq.~\eqref{eq:rhoinfgamma} is a power decay of $\rho_{\infty}(t)$ at large times that is clearly seen in the double logarithmic scale in the inset of Fig.~\ref{figure4}(a). Instead of the exponential decay, expected for lossy systems, the time response function has the asymptotics $\rho_{\infty}(t) \approx \sqrt{\Gamma/(\pi\Delta^2 t^3)}$ at $t \gg 1/\Gamma$. Such a behavior is closely related to the Borrmann effect, known in the $X$-ray physics as the origin of extraordinary transmission of radiation through  absorptive media.\cite{borrmann1950} In our case of resonant Bragg structures, the effect comes from the suppression of radiation absorption at the resonant frequency $\omega_0$. Indeed, in the semi-infinite structure, the steady-state distribution of radiation electric field at the resonance frequency is a standing wave with the nodes situated at the QW centers.\cite{Ivchenko1994} Such a wave does not interact with excitons in QWs. This leads to a narrow dip in the absorption spectrum and a narrow peak in the reflection spectrum $|r(\omega)|^2$ at the frequency $\omega_0$ with the characteristic width of $\Gamma$,\cite{SpecFreqs2} see Fig.~\ref{figure4}(b) and the inset. In the semi-infinite QW structure, $|r(\omega_0)|=1$, which results in a power decay of $\rho_{\infty}(t)$ at large times.

In a long but finite structure, the peak in the reflection spectrum remains, however, $|r_N(\omega)|$ does not reach unity, see the inset in Fig.~\ref{figure4}(b). Therefore, a part of the response function with a power-like decay can be still distinguished but the behavior of $\rho_N(t)$ at $t \rightarrow \infty$ is always described by the exponent. Such a response function is shown in Fig.~\ref{figure4}(a) by red solid curve. The decay rate of $\rho_N(t)$ is determined by the pole of Eq.~(\ref{eq:pnw}) with the smallest negative imaginary part, which yields the asymptotics $\rho_N(t) \approx \pi^2 \Gamma^2 /(2 \nu^3 \Delta) \exp(-\pi^2 \Gamma t/ 2\nu)$ at $t \gg \nu^3 / \Gamma$.

\section{Aperiodic Bragg structures}\label{sec:abs}

In this section, we analyze the effects of non-periodicity of QW arrangement and detuning from the Bragg condition on the optical response. We focus on quasiperiodic Fibonacci structures (Sec.~\ref{ssec:fs}) and gradient structures where the interwell distance gradually changes from the structure forefront to its back edge (Sec.~\ref{ssec:gs}).    

\subsection{Fibonacci structures}\label{ssec:fs}

In canonic Fibonacci QW chains, the interwell distances 
can take one of the two values, $a$ or $b$, 
with $a/b$ being equal to the golden ratio $\tau=(\sqrt{5}+1)/2\approx 1.62$. The sequence of the distances is determined by the recurrent relation similar to that for the Fibonacci numbers. The sequence of the  order $m$, $\mathcal{F}_m$, is given by $\mathcal{F}_{m-1} \mathcal{F}_{m-2}$ with $\mathcal{F}_1=b$ and $\mathcal{F}_2=a$.\cite{Janot} In particular, the chain $\mathcal{F}_5=a b a a b$ describes the structure consisting of $N=6$ quantum wells positioned at $z_1=0$, $z_2=a$, $z_3=a+b$, $z_4=2a+b$, $z_5=3a+b$, and $z_6=3a+2b$. 

Despite the non-periodicity, the Fibonacci QW structures can be tuned to the resonant reflection.\cite{Poddubny2008prb}
The corresponding Bragg condition for such structures has the form
\begin{equation}\label{eq:fibbrag}
\frac{\omega_0 n_b}{c} \bar{d} = \pi \left( h + \frac{h'}{\tau} \right) \:,
\end{equation}
where $\bar{d}=(3-\tau)b$ is the mean interwell distance and $h$ and $h'$ are the integer numbers. In contrast to periodic systems, the Bragg resonances in the Fibonacci structures are characterized by two numbers $h$ and $h'$, which reflects the fact that these structures can be obtained by the cut-and-project method from the two-dimensional square lattice.\cite{Janot,cryst2006}

\begin{figure}[t!]
\includegraphics[width=0.49\textwidth]{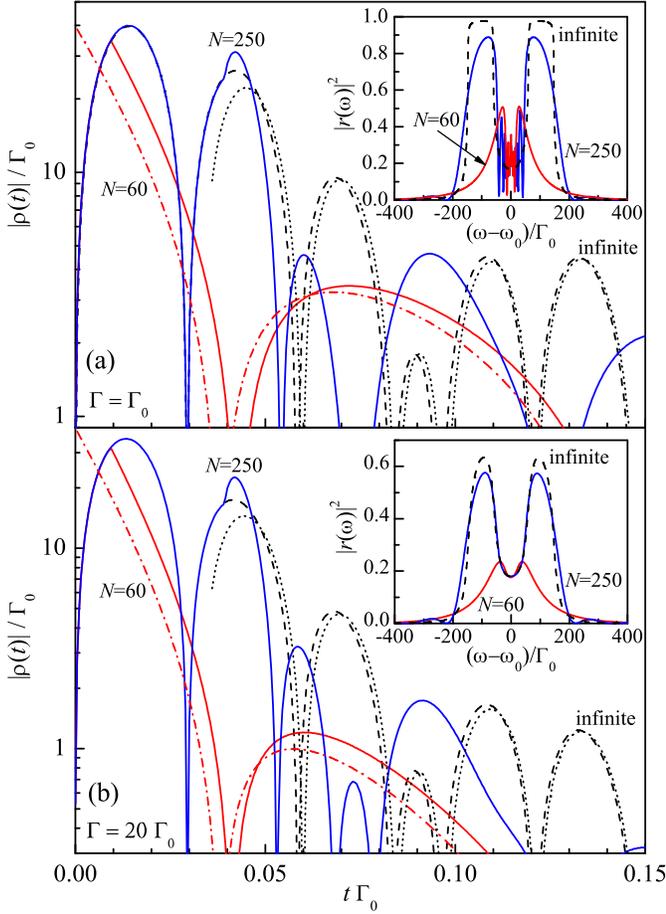}
\caption{(Color online) The envelopes of reflected pulses calculated for the Fibonacci structures with various numbers of QWs, the Bragg condition $(h,h')=(1,0)$, and $\omega_0/\Gamma_0 = 4 \times 10^4$. Figures~(a) and~(b) correspond to weak, $\Gamma=\Gamma_0$, and strong, $\Gamma=20 \Gamma_0$, non-radiative decay, respectively. Black dashed curves depict the pulse envelopes for the semi-infinite Fibonacci structure. Black dotted curves present the analytical asymptotics Eq.~(\ref{rho_Fib_inf}). Red dash-dotted curves are plotted after analytical Eq.~(\ref{eq:bessel}) obtained for the superradiant regime. The insets show the reflection spectra of the same QW structures in the frequency domain.}
\label{figure5}
\end{figure}

The envelopes of reflected pulses in the time domain for the Fibonacci structures tuned to the Bragg resonance $(h,h')=(1,0)$ are shown in Fig.~\ref{figure5} by solid curves. Figures~\ref{figure5}(a) and~\ref{figure5}(b) correspond to the cases of weak and strong non-radiative decay, respectively. The curves are calculated numerically by the transfer matrix technique. 
One can see that the reflected pulses exhibit oscillations even for structures with rather small QW numbers $N$ which operate in the superradiant regime. This is in contrast to periodic Bragg structures where the oscillations emerge in the photonic-crystalline regime, i.e., at large $N$, see Fig.~\ref{figure2}. The origin of such a behavior is a dip in the
reflection spectrum of the Fibonacci structures at $\omega \approx \omega_0$, see insets in Figs.~\ref{figure5}(a) and~\ref{figure5}(b), which is caused by non-periodicity of the QW arrangement.\cite{Poddubny2008prb} Similar spectrum structure has been observed for periodic Bragg systems with two QWs in the unit cell\cite{Voronov2004} and for layered nuclear systems.\cite{chumakov1993}

To analyze the reflection of optical pulses from the Fibonacci structures in more detail we use the two-wave approximation. In this approach, the eigen electromagnetic states of the structure in the vicinity of the Bragg resonance $(h,h')$ are approximated by the sum of two plane waves with the wave vectors $K$ and $K-G_{hh'}$, with
$G_{hh'}=(2\pi/\bar{d})(h+h'/\tau)$ being the diffraction vector.\cite{poddubny2009}
The approach yields the reflection spectra of the Fibonacci structures
\begin{equation}\label{eq:26}
r_N(\omega) = \frac{\chi \f}{\omega_0 - \omega -\chi - \rmi \Omega\cot(2\Omega N \Gamma_0/\Delta^2_{hh'})} \:,
\end{equation}
where
\[
\chi = \frac{\Delta^2_{hh'}}{2(\omega_0-\omega-\rmi\Gamma)} \:,\;\; \Omega = \sqrt{(\chi+\omega-\omega_0)^2-\chi^2 |\f|^2} \:,
\]
$\Delta_{hh'}=\Delta/\sqrt{h+h'/\tau}$, and $f_{hh'}$ is the structure factor\cite{Janot} determining the strength of the Bragg diffraction peak,
\begin{equation}\label{f_factor}
f_{hh'} = \frac{\sin S_{hh'}}{S_{hh'}} \exp\Bigl({\rm i}
\frac{\tau-2}{\tau} S_{hh'}\Bigr) \:, \;\; S_{hh'} =\frac{\pi
\tau (\tau h'-h)}{\tau^2+1} \:.
\end{equation}
In a simple periodic structure $f_{hh'}=1$ and Eq.~\eqref{eq:26} reduces to Eq.~\eqref{eq:pnw}, while in quasicrystals $|f_{hh'}|<1$. Comparison of the reflection spectra given by Eq.~\eqref{eq:26} with those obtained numerically by the transfer matrix technique shows that Eq.~\eqref{eq:26} is valid in the whole frequency range of the resonant reflection 
except for a narrow central region.

In the regime of superradiance, 
Eq.~\eqref{eq:26} assumes the form
\begin{equation}\label{eq:26sr}
r_N (\omega) = -\frac{\f}{1+\rmi\alpha_{hh'}\cot\left(\dfrac{\alpha_{hh'} N\Gamma_0}{\omega_0-\omega-\rmi\Gamma}\right)} \:,
\end{equation}
where $\alpha_{hh'} = \sqrt{1-|\f|^2}$ characterizes the strength of aperiodicity. For a periodic structure $\alpha_{hh'}=0$ and Eq.~\eqref{eq:26sr} reduces to Eq.~\eqref{eq:rsr}. The reflection spectrum given by Eq.~\eqref{eq:26sr} has a wide peak at the frequency $\omega_0$ with the width proportional to $N\Gamma_0$, which is a signature of the superradiant regime. In the middle of the peak, there is a structured dip of the width $\sim \alpha_{hh'}N \Gamma_0$, which is caused by the spacial nonperiodicity of the system. Such a multipeak structure of the reflection spectrum $r_N(\omega)$ leads, in turn, to a beating pattern in the response function $\rho_N(t)$ in the time domain, see also Ref.~\onlinecite{chumakov1999}. To obtain analytical expression for $\rho_N(t)$, we expand Eq.~\eqref{eq:26sr} in the Taylor series at $\alpha_{hh'} \rightarrow 1$ and perform the Fourier transformation Eq.~(\ref{eq:rho}). This procedure yields
\begin{multline}\label{eq:bessel}
\rho_N(t) = - 2 \f  \sum_{k=1}^{\infty} \frac{\alpha_{hh'} (1-\alpha_{hh'})^{k-1}}{(1+\alpha_{hh'})^{k+1}} \sqrt{\frac{2 k \alpha_{hh'} N\Gamma_0}{t}} \\ \times J_1(\sqrt{8k\alpha_{hh'} N\Gamma_0 t})\,{\rm e}^{-\Gamma t} {\rm e}^{-\rmi \omega_0 t} \theta(t) \:.
\end{multline}
The response functions given by Eq.~(\ref{eq:bessel}) for the QW number $N=60$ and different non-radiative decay rates are plotted by dash-dotted curves in Figs.~\ref{figure5}(a) and~\ref{figure5}(b). One can see that the equation describes the exact solutions shown by solid curves quite well. We note that for small numbers of $h$ and $h'$, where the parameter $\alpha_{hh'}$ is close to unity, the dominant contribution to $\rho_N(t)$ comes from the first term ($k=1$) of the series~\eqref{eq:bessel}.

In the opposite case of very long structures, $N \gg \sqrt{\omega_o/\Gamma_0}$, the multiple reflection of light from quantum wells in the Fibonacci structure leads to the formation of two wide stop bands, see insets in Fig.~\ref{figure5}(a) and~\ref{figure5}(b). The corresponding reflection coefficient $r_{\infty}(\omega)$ for the semi-infinite structure is obtained from Eq.~(\ref{eq:26}) by the replacement of $\cot(2\Omega N \Gamma_0/\Delta^2_{hh'})$ in the denominator with $(-{\rm i}) {\rm sign} ({\rm Im} \Omega)$. The subsequent Fourier transformation of $r_{\infty}(\omega)$ by the stationary phase method yields the response function in the time domain
\newcommand{\D}{\Delta_{h,h'}}
\begin{multline}\label{rho_Fib_inf}
\rho_{\infty}(t) \approx \f \sqrt{\frac{8}{\pi |\f|^3 \D^2 t^3}} \, \e^{-\Gamma t /2} \eiwt \theta(t) \\
\times \left[ \sqrt{\omega_+} \sin \left( \omega_+ t + \frac{\pi}{4} \right) - \sqrt{\omega_-} \cos \left( \omega_- t + \frac{\pi}{4} \right) \right] \:,
\end{multline}
which is valid at $t \gg |\omega_{\pm}|$. Here, the frequencies $\omega_{\pm} = \sqrt{ \D^2 (1 \pm |\f|)/2 - \Gamma^2/4 }$ determine the inner and outer edges of the stop bands measured from $\omega_0$. The analytical function~(\ref{rho_Fib_inf}) shown in Figs.~\ref{figure5}(a) and~\ref{figure5}(b) by dotted curves is in a perfect agreement with the results of numerical calculation by the transfer matrix technique (dashed curves) at large times. 

The transition from superradiant to photonic-crystalline regime with the increase of the QW number leads to a change in the normalized energy $\mathcal{P}_N$ and duration $\mathcal{T}_N$ of the reflected pulse, Eq.~(\ref{eq:PT}).
The dependences of $\mathcal{P}_N$ and $\mathcal{T}_N$ on the number of QWs for the Fibonacci structures with different indices $(h,h')$ are shown in Fig.~\ref{figure3}(a) and~\ref{figure3}(b), respectively. They are qualitatively similar to those for periodic structures. The energy $\mathcal{P}_N$ increases linearly with the QW number at small $N$ and saturates at $N\to\infty$. The duration $\mathcal{T}_N$ has a minimum at $N\sim\sqrt{\omega_0/\Gamma_0}$, where the transition from superradiant to photonic-crystalline regime occurs. The dependence of $\mathcal{P}_N$ and $\mathcal{T}_N$ on the indices  $(h,h')$ is determined by two competing effects. First, the structure factor $|f_{hh'}|$ grows with $(h,h')$ [see Eq.~(\ref{f_factor})], which enhances the structure reflectivity. Second, the average period $\bar{d}$ increases with $(h,h')$ [Eq.~\eqref{eq:fibbrag}] leading to the increase in the flight time and, by that, to the suppression of the reflectivity. At large numbers of $(h,h')$, when the structure factor is saturated and close to unity, the second effect dominates. As a result, the energy of the reflected pulse $\mathcal{P}_N$ becomes smaller while the pulse duration $\mathcal{T}_N$ becomes longer. This explains the difference between the curves in Fig.~\ref{figure3}.

\subsection{Gradient structures}\label{ssec:gs}

In addition to periodic or quasiperiodic structures, which may be characterized by the mean period, one can consider
spatially inhomogeneous photonic systems. As an example of such systems, we analyze here the gradient structures, where the distance between the neighboring QWs gradually changes from the structure forefront to its back edge. We assume that the positions of QWs are given by
\begin{equation}\label{eq:gradient}
z_j=(j-1)(1+\zeta/j)d \:,
\end{equation} 
where $\zeta$ is a dimensional parameter describing the structure inhomogeneity and $d$ is the length given by the Bragg reflection condition~\eqref{eq:bragg}. The front part of the structure is substantially detuned from the Bragg condition while at $j \gg \zeta$
the interwell distance tends to $d$ and the structure transforms to the periodic Bragg structure. The optical response of the gradient structure is determi\bibliography{all_cp1251b}ned by the parameter $\zeta$, the QW number $N$, as well as the radiative and non-radiative decay rates $\Gamma_0$ and $\Gamma$, respectively. 

\begin{thebibliography}{39}%
\makeatletter
\providecommand \@ifxundefined [1]{%
 \@ifx{#1\undefined}
}%
\providecommand \@ifnum [1]{%
 \ifnum #1\expandafter \@firstoftwo
 \else \expandafter \@secondoftwo
 \fi
}%
\providecommand \@ifx [1]{%
 \ifx #1\expandafter \@firstoftwo
 \else \expandafter \@secondoftwo
 \fi
}%
\providecommand \natexlab [1]{#1}%
\providecommand \enquote  [1]{``#1''}%
\providecommand \bibnamefont  [1]{#1}%
\providecommand \bibfnamefont [1]{#1}%
\providecommand \citenamefont [1]{#1}%
\providecommand \href@noop [0]{\@secondoftwo}%
\providecommand \href [0]{\begingroup \@sanitize@url \@href}%
\providecommand \@href[1]{\@@startlink{#1}\@@href}%
\providecommand \@@href[1]{\endgroup#1\@@endlink}%
\providecommand \@sanitize@url [0]{\catcode `\\12\catcode `\$12\catcode
  `\&12\catcode `\#12\catcode `\^12\catcode `\_12\catcode `\%12\relax}%
\providecommand \@@startlink[1]{}%
\providecommand \@@endlink[0]{}%
\providecommand \url  [0]{\begingroup\@sanitize@url \@url }%
\providecommand \@url [1]{\endgroup\@href {#1}{\urlprefix }}%
\providecommand \urlprefix  [0]{URL }%
\providecommand \Eprint [0]{\href }%
\providecommand \doibase [0]{http://dx.doi.org/}%
\providecommand \selectlanguage [0]{\@gobble}%
\providecommand \bibinfo  [0]{\@secondoftwo}%
\providecommand \bibfield  [0]{\@secondoftwo}%
\providecommand \translation [1]{[#1]}%
\providecommand \BibitemOpen [0]{}%
\providecommand \bibitemStop [0]{}%
\providecommand \bibitemNoStop [0]{.\EOS\space}%
\providecommand \EOS [0]{\spacefactor3000\relax}%
\providecommand \BibitemShut  [1]{\csname bibitem#1\endcsname}%
\let\auto@bib@innerbib\@empty
\bibitem [{\citenamefont {Limonov}\ and\ \citenamefont {{M. De La
  Rue}}(2012)}]{Limonov_book}%
  \BibitemOpen
  \bibinfo {editor} {\bibfnamefont {Mikhail~F.}\ \bibnamefont {Limonov}}\ and\
  \bibinfo {editor} {\bibfnamefont {Richard}\ \bibnamefont {{M. De La Rue}}},\
  eds.,\ \href@noop {} {\emph {\bibinfo {title} {{O}ptical {P}roperties {O}f
  {P}hotonic {S}tructures: {I}nterplay {O}f {O}rder {A}nd {D}isorder}}}\
  (\bibinfo  {publisher} {Taylor \& Francis},\ \bibinfo {address} {New York},\
  \bibinfo {year} {2012})\BibitemShut {NoStop}%
\bibitem [{\citenamefont {Ivchenko}(2005)}]{Ivchenko_book}%
  \BibitemOpen
  \bibfield  {author} {\bibinfo {author} {\bibfnamefont {E.~L.}\ \bibnamefont
  {Ivchenko}},\ }\href@noop {} {\emph {\bibinfo {title} {{O}ptical spectroscopy
  of semiconductor nanostructures}}}\ (\bibinfo  {publisher} {Alpha Science
  International},\ \bibinfo {address} {Harrow, UK},\ \bibinfo {year}
  {2005})\BibitemShut {NoStop}%
\bibitem [{\citenamefont {Kagan}(1999)}]{kagan1999}%
  \BibitemOpen
  \bibfield  {author} {\bibinfo {author} {\bibfnamefont {Yu}~\bibnamefont
  {Kagan}},\ }\bibfield  {title} {\enquote {\bibinfo {title} {{T}heory of
  coherent phenomena and fundamentals in nuclear resonant scattering},}\
  }\href@noop {} {\bibfield  {journal} {\bibinfo  {journal} {Hyperfine
  Interactions}\ }\textbf {\bibinfo {volume} {123}},\ \bibinfo {pages}
  {83--126} (\bibinfo {year} {1999})}\BibitemShut {NoStop}%
\bibitem [{\citenamefont {{Shvyd'ko}}\ \emph {et~al.}(2010)\citenamefont
  {{Shvyd'ko}}, \citenamefont {{Stoupin}}, \citenamefont {{Cunsolo}},
  \citenamefont {{Said}},\ and\ \citenamefont {{Huang}}}]{Shvydko2010}%
  \BibitemOpen
  \bibfield  {author} {\bibinfo {author} {\bibfnamefont {Y.~V.}\ \bibnamefont
  {{Shvyd'ko}}}, \bibinfo {author} {\bibfnamefont {S.}~\bibnamefont
  {{Stoupin}}}, \bibinfo {author} {\bibfnamefont {A.}~\bibnamefont
  {{Cunsolo}}}, \bibinfo {author} {\bibfnamefont {A.~H.}\ \bibnamefont
  {{Said}}}, \ and\ \bibinfo {author} {\bibfnamefont {X.}~\bibnamefont
  {{Huang}}},\ }\bibfield  {title} {\enquote {\bibinfo {title}
  {{High-reflectivity high-resolution X-ray crystal optics with diamonds}},}\
  }\href {\doibase 10.1038/nphys1506} {\bibfield  {journal} {\bibinfo
  {journal} {Nat. Physics}\ }\textbf {\bibinfo {volume} {6}},\ \bibinfo {pages}
  {196--199} (\bibinfo {year} {2010})}\BibitemShut {NoStop}%
\bibitem [{\citenamefont {Feltin}\ \emph {et~al.}(2006)\citenamefont {Feltin},
  \citenamefont {Carlin}, \citenamefont {Dorsaz}, \citenamefont {Christmann},
  \citenamefont {Butte}, \citenamefont {Laugt}, \citenamefont {Ilegems},\ and\
  \citenamefont {Grandjean}}]{Feltin2006}%
  \BibitemOpen
  \bibfield  {author} {\bibinfo {author} {\bibfnamefont {E.}~\bibnamefont
  {Feltin}}, \bibinfo {author} {\bibfnamefont {J.-F.}\ \bibnamefont {Carlin}},
  \bibinfo {author} {\bibfnamefont {J.}~\bibnamefont {Dorsaz}}, \bibinfo
  {author} {\bibfnamefont {G.}~\bibnamefont {Christmann}}, \bibinfo {author}
  {\bibfnamefont {R.}~\bibnamefont {Butte}}, \bibinfo {author} {\bibfnamefont
  {M.}~\bibnamefont {Laugt}}, \bibinfo {author} {\bibfnamefont
  {M.}~\bibnamefont {Ilegems}}, \ and\ \bibinfo {author} {\bibfnamefont
  {N.}~\bibnamefont {Grandjean}},\ }\bibfield  {title} {\enquote {\bibinfo
  {title} {{C}rack-free highly reflective $\mbox{AlInN/AlGaN}$ $\mbox{B}$ragg
  mirrors for $\mbox{UV}$ applications},}\ }\href {\doibase 10.1063/1.2167399}
  {\bibfield  {journal} {\bibinfo  {journal} {Appl. Phys. Lett.}\ }\textbf
  {\bibinfo {volume} {88}},\ \bibinfo {eid} {051108} (\bibinfo {year}
  {2006})}\BibitemShut {NoStop}%
\bibitem [{\citenamefont {{Wierer}}\ \emph {et~al.}(2009)\citenamefont
  {{Wierer}}, \citenamefont {{David}},\ and\ \citenamefont
  {{Megens}}}]{Megens2009}%
  \BibitemOpen
  \bibfield  {author} {\bibinfo {author} {\bibfnamefont {J.~J.}\ \bibnamefont
  {{Wierer}}}, \bibinfo {author} {\bibfnamefont {A.}~\bibnamefont {{David}}}, \
  and\ \bibinfo {author} {\bibfnamefont {M.~M.}\ \bibnamefont {{Megens}}},\
  }\bibfield  {title} {\enquote {\bibinfo {title} {{III-nitride
  photonic-crystal light-emitting diodes with high extraction efficiency}},}\
  }\href {\doibase 10.1038/nphoton.2009.21} {\bibfield  {journal} {\bibinfo
  {journal} {Nat. Photonics}\ }\textbf {\bibinfo {volume} {3}},\ \bibinfo
  {pages} {163--169} (\bibinfo {year} {2009})}\BibitemShut {NoStop}%
\bibitem [{\citenamefont {Deutsch}\ \emph {et~al.}(1995)\citenamefont
  {Deutsch}, \citenamefont {Spreeuw}, \citenamefont {Rolston},\ and\
  \citenamefont {Phillips}}]{deutsch1995}%
  \BibitemOpen
  \bibfield  {author} {\bibinfo {author} {\bibfnamefont {I.~H.}\ \bibnamefont
  {Deutsch}}, \bibinfo {author} {\bibfnamefont {R.~J.~C.}\ \bibnamefont
  {Spreeuw}}, \bibinfo {author} {\bibfnamefont {S.~L.}\ \bibnamefont
  {Rolston}}, \ and\ \bibinfo {author} {\bibfnamefont {W.~D.}\ \bibnamefont
  {Phillips}},\ }\bibfield  {title} {\enquote {\bibinfo {title} {{P}hotonic
  band gaps in optical lattices},}\ }\href {\doibase 10.1103/PhysRevA.52.1394}
  {\bibfield  {journal} {\bibinfo  {journal} {Phys. Rev. A}\ }\textbf {\bibinfo
  {volume} {52}},\ \bibinfo {pages} {1394--1410} (\bibinfo {year}
  {1995})}\BibitemShut {NoStop}%
\bibitem [{\citenamefont {Yanik}\ \emph {et~al.}(2004)\citenamefont {Yanik},
  \citenamefont {Suh}, \citenamefont {Wang},\ and\ \citenamefont
  {Fan}}]{yanik2004}%
  \BibitemOpen
  \bibfield  {author} {\bibinfo {author} {\bibfnamefont {Mehmet~Fatih}\
  \bibnamefont {Yanik}}, \bibinfo {author} {\bibfnamefont {Wonjoo}\
  \bibnamefont {Suh}}, \bibinfo {author} {\bibfnamefont {Zheng}\ \bibnamefont
  {Wang}}, \ and\ \bibinfo {author} {\bibfnamefont {Shanhui}\ \bibnamefont
  {Fan}},\ }\bibfield  {title} {\enquote {\bibinfo {title} {{S}topping {L}ight
  in a {W}aveguide with an {A}ll-{O}ptical {A}nalog of {E}lectromagnetically
  {I}nduced {T}ransparency},}\ }\href {\doibase 10.1103/PhysRevLett.93.233903}
  {\bibfield  {journal} {\bibinfo  {journal} {Phys. Rev. Lett.}\ }\textbf
  {\bibinfo {volume} {93}},\ \bibinfo {pages} {233903} (\bibinfo {year}
  {2004})}\BibitemShut {NoStop}%
\bibitem [{\citenamefont {Kavokin}\ \emph {et~al.}(2006)\citenamefont
  {Kavokin}, \citenamefont {Baumberg}, \citenamefont {Malpuech},\ and\
  \citenamefont {Laussy}}]{Kavbamalas_book}%
  \BibitemOpen
  \bibfield  {author} {\bibinfo {author} {\bibfnamefont {A.}~\bibnamefont
  {Kavokin}}, \bibinfo {author} {\bibfnamefont {J.J.}\ \bibnamefont
  {Baumberg}}, \bibinfo {author} {\bibfnamefont {G.}~\bibnamefont {Malpuech}},
  \ and\ \bibinfo {author} {\bibfnamefont {F.P.}\ \bibnamefont {Laussy}},\
  }\href@noop {} {\emph {\bibinfo {title} {{M}icrocavities}}}\ (\bibinfo
  {publisher} {Clarendon Press},\ \bibinfo {address} {Oxford},\ \bibinfo {year}
  {2006})\BibitemShut {NoStop}%
\bibitem [{\citenamefont {Prineas}\ \emph {et~al.}(2006)\citenamefont
  {Prineas}, \citenamefont {Johnston}, \citenamefont {Yildirim}, \citenamefont
  {Zhao},\ and\ \citenamefont {Smirl}}]{prineas2006apl}%
  \BibitemOpen
  \bibfield  {author} {\bibinfo {author} {\bibfnamefont {J.~P.}\ \bibnamefont
  {Prineas}}, \bibinfo {author} {\bibfnamefont {W.~J.}\ \bibnamefont
  {Johnston}}, \bibinfo {author} {\bibfnamefont {M.}~\bibnamefont {Yildirim}},
  \bibinfo {author} {\bibfnamefont {J.}~\bibnamefont {Zhao}}, \ and\ \bibinfo
  {author} {\bibfnamefont {Arthur~L.}\ \bibnamefont {Smirl}},\ }\bibfield
  {title} {\enquote {\bibinfo {title} {{T}unable slow light in {B}ragg-spaced
  quantum wells},}\ }\href {\doibase 10.1063/1.2403927} {\bibfield  {journal}
  {\bibinfo  {journal} {Appl. Phys. Lett.}\ }\textbf {\bibinfo {volume} {89}},\
  \bibinfo {eid} {241106} (\bibinfo {year} {2006})}\BibitemShut {NoStop}%
\bibitem [{\citenamefont {Chaldyshev}\ \emph
  {et~al.}(2011{\natexlab{a}})\citenamefont {Chaldyshev}, \citenamefont {Chen},
  \citenamefont {Poddubny}, \citenamefont {Vasil'ev},\ and\ \citenamefont
  {Liu}}]{chaldyshev2011}%
  \BibitemOpen
  \bibfield  {author} {\bibinfo {author} {\bibfnamefont {V.~V.}\ \bibnamefont
  {Chaldyshev}}, \bibinfo {author} {\bibfnamefont {Yuechao}\ \bibnamefont
  {Chen}}, \bibinfo {author} {\bibfnamefont {A.~N.}\ \bibnamefont {Poddubny}},
  \bibinfo {author} {\bibfnamefont {A.~P.}\ \bibnamefont {Vasil'ev}}, \ and\
  \bibinfo {author} {\bibfnamefont {Zhiheng}\ \bibnamefont {Liu}},\ }\bibfield
  {title} {\enquote {\bibinfo {title} {{R}esonant optical reflection by a
  periodic system of the quantum well excitons at the second quantum state},}\
  }\href {\doibase 10.1063/1.3554429} {\bibfield  {journal} {\bibinfo
  {journal} {Appl. Phys. Lett.}\ }\textbf {\bibinfo {volume} {98}},\ \bibinfo
  {eid} {073112} (\bibinfo {year} {2011}{\natexlab{a}})}\BibitemShut {NoStop}%
\bibitem [{\citenamefont {{Ivchenko}}\ \emph {et~al.}(1994)\citenamefont
  {{Ivchenko}}, \citenamefont {{Nesvizhskii}},\ and\ \citenamefont
  {{Jorda}}}]{Ivchenko1994}%
  \BibitemOpen
  \bibfield  {author} {\bibinfo {author} {\bibfnamefont {E.~L.}\ \bibnamefont
  {{Ivchenko}}}, \bibinfo {author} {\bibfnamefont {A.~I.}\ \bibnamefont
  {{Nesvizhskii}}}, \ and\ \bibinfo {author} {\bibfnamefont {S.}~\bibnamefont
  {{Jorda}}},\ }\bibfield  {title} {\enquote {\bibinfo {title} {{B}ragg
  reflection of light from quantum-well structures},}\ }\href@noop {}
  {\bibfield  {journal} {\bibinfo  {journal} {Fiz. Tverd. Tela}\ }\textbf
  {\bibinfo {volume} {36}},\ \bibinfo {pages} {2118} (\bibinfo {year}
  {1994})},\ \translation{Phys. Solid State {\bf 36}, 1156 (1994)}\BibitemShut
  {NoStop}%
\bibitem [{\citenamefont {H{\"u}bner}\ \emph {et~al.}(1996)\citenamefont
  {H{\"u}bner}, \citenamefont {Kuhl}, \citenamefont {Stroucken}, \citenamefont
  {Knorr}, \citenamefont {Koch}, \citenamefont {Hey},\ and\ \citenamefont
  {Ploog}}]{koch1996mqw}%
  \BibitemOpen
  \bibfield  {author} {\bibinfo {author} {\bibfnamefont {M.}~\bibnamefont
  {H{\"u}bner}}, \bibinfo {author} {\bibfnamefont {J.}~\bibnamefont {Kuhl}},
  \bibinfo {author} {\bibfnamefont {T.}~\bibnamefont {Stroucken}}, \bibinfo
  {author} {\bibfnamefont {A.}~\bibnamefont {Knorr}}, \bibinfo {author}
  {\bibfnamefont {S.~W.}\ \bibnamefont {Koch}}, \bibinfo {author}
  {\bibfnamefont {R.}~\bibnamefont {Hey}}, \ and\ \bibinfo {author}
  {\bibfnamefont {K.}~\bibnamefont {Ploog}},\ }\bibfield  {title} {\enquote
  {\bibinfo {title} {{C}ollective {E}ffects of {E}xcitons in
  {M}ultiple-{Q}uantum-{W}ell {B}ragg and {A}nti-{B}ragg {S}tructures},}\
  }\href {\doibase 10.1103/PhysRevLett.76.4199} {\bibfield  {journal} {\bibinfo
   {journal} {Phys. Rev. Lett.}\ }\textbf {\bibinfo {volume} {76}},\ \bibinfo
  {pages} {4199--4202} (\bibinfo {year} {1996})}\BibitemShut {NoStop}%
\bibitem [{\citenamefont {Ikawa}\ and\ \citenamefont {Cho}(2002)}]{Cho2002}%
  \BibitemOpen
  \bibfield  {author} {\bibinfo {author} {\bibfnamefont {Tomoe}\ \bibnamefont
  {Ikawa}}\ and\ \bibinfo {author} {\bibfnamefont {Kikuo}\ \bibnamefont
  {Cho}},\ }\bibfield  {title} {\enquote {\bibinfo {title} {{F}ate of the
  superradiant mode in a resonant $\mbox{B}$ragg reflector},}\ }\href {\doibase
  10.1103/PhysRevB.66.085338} {\bibfield  {journal} {\bibinfo  {journal} {Phys.
  Rev. B}\ }\textbf {\bibinfo {volume} {66}},\ \bibinfo {pages} {085338}
  (\bibinfo {year} {2002})}\BibitemShut {NoStop}%
\bibitem [{\citenamefont {Pilozzi}\ \emph {et~al.}(2004)\citenamefont
  {Pilozzi}, \citenamefont {D'Andrea},\ and\ \citenamefont
  {Cho}}]{Pilozzi2004}%
  \BibitemOpen
  \bibfield  {author} {\bibinfo {author} {\bibfnamefont {L.}~\bibnamefont
  {Pilozzi}}, \bibinfo {author} {\bibfnamefont {A.}~\bibnamefont {D'Andrea}}, \
  and\ \bibinfo {author} {\bibfnamefont {K.}~\bibnamefont {Cho}},\ }\bibfield
  {title} {\enquote {\bibinfo {title} {{S}patial dispersion effects on the
  optical properties of a resonant $\mbox{B}$ragg reflector},}\ }\href
  {\doibase 10.1103/PhysRevB.69.205311} {\bibfield  {journal} {\bibinfo
  {journal} {Phys. Rev. B}\ }\textbf {\bibinfo {volume} {69}},\ \bibinfo {eid}
  {205311} (\bibinfo {year} {2004})}\BibitemShut {NoStop}%
\bibitem [{\citenamefont {Ivchenko}\ \emph {et~al.}(2004)\citenamefont
  {Ivchenko}, \citenamefont {Voronov}, \citenamefont {Erementchouk},
  \citenamefont {Deych},\ and\ \citenamefont {Lisyansky}}]{Voronov2004}%
  \BibitemOpen
  \bibfield  {author} {\bibinfo {author} {\bibfnamefont {E.~L.}\ \bibnamefont
  {Ivchenko}}, \bibinfo {author} {\bibfnamefont {M.~M.}\ \bibnamefont
  {Voronov}}, \bibinfo {author} {\bibfnamefont {M.~V.}\ \bibnamefont
  {Erementchouk}}, \bibinfo {author} {\bibfnamefont {L.~I.}\ \bibnamefont
  {Deych}}, \ and\ \bibinfo {author} {\bibfnamefont {A.~A.}\ \bibnamefont
  {Lisyansky}},\ }\bibfield  {title} {\enquote {\bibinfo {title}
  {{M}ultiple-quantum-well-based photonic crystals with simple and compound
  elementary supercells},}\ }\href {\doibase 10.1103/PhysRevB.70.195106}
  {\bibfield  {journal} {\bibinfo  {journal} {Phys. Rev. B}\ }\textbf {\bibinfo
  {volume} {70}},\ \bibinfo {pages} {195106} (\bibinfo {year}
  {2004})}\BibitemShut {NoStop}%
\bibitem [{\citenamefont {Averkiev}\ \emph {et~al.}(2012)\citenamefont
  {Averkiev}, \citenamefont {Glazov},\ and\ \citenamefont
  {Voronov}}]{Averkiev2012}%
  \BibitemOpen
  \bibfield  {author} {\bibinfo {author} {\bibfnamefont {N.S.}\ \bibnamefont
  {Averkiev}}, \bibinfo {author} {\bibfnamefont {M.M.}\ \bibnamefont {Glazov}},
  \ and\ \bibinfo {author} {\bibfnamefont {M.M.}\ \bibnamefont {Voronov}},\
  }\bibfield  {title} {\enquote {\bibinfo {title} {{F}ermi-edge polaritons in
  {B}ragg multiple-quantum-well structures},}\ }\href {\doibase
  10.1016/j.ssc.2011.12.002} {\bibfield  {journal} {\bibinfo  {journal} {Solid
  State Comm.}\ }\textbf {\bibinfo {volume} {152}},\ \bibinfo {pages} {395 --
  398} (\bibinfo {year} {2012})}\BibitemShut {NoStop}%
\bibitem [{\citenamefont {{Kochereshko}}\ \emph {et~al.}(1994)\citenamefont
  {{Kochereshko}}, \citenamefont {{Pozina}}, \citenamefont {{Ivchenko}},
  \citenamefont {{Yakovlev}}, \citenamefont {{Waag}}, \citenamefont {{Ossau}},
  \citenamefont {{Landwehr}}, \citenamefont {{Hellmann}},\ and\ \citenamefont
  {{G{\"o}bel}}}]{Kochereshko1994}%
  \BibitemOpen
  \bibfield  {author} {\bibinfo {author} {\bibfnamefont {V.~P.}\ \bibnamefont
  {{Kochereshko}}}, \bibinfo {author} {\bibfnamefont {G.~R.}\ \bibnamefont
  {{Pozina}}}, \bibinfo {author} {\bibfnamefont {E.~L.}\ \bibnamefont
  {{Ivchenko}}}, \bibinfo {author} {\bibfnamefont {D.~R.}\ \bibnamefont
  {{Yakovlev}}}, \bibinfo {author} {\bibfnamefont {A.}~\bibnamefont {{Waag}}},
  \bibinfo {author} {\bibfnamefont {W.}~\bibnamefont {{Ossau}}}, \bibinfo
  {author} {\bibfnamefont {G.}~\bibnamefont {{Landwehr}}}, \bibinfo {author}
  {\bibfnamefont {R.}~\bibnamefont {{Hellmann}}}, \ and\ \bibinfo {author}
  {\bibfnamefont {E.~O.}\ \bibnamefont {{G{\"o}bel}}},\ }\bibfield  {title}
  {\enquote {\bibinfo {title} {{G}iant exciton resonance reflectance in
  $\mbox{B}$ragg $\mbox{MQW}$ structures},}\ }\href {\doibase
  10.1006/spmi.1994.1090} {\bibfield  {journal} {\bibinfo  {journal}
  {Superlatt. Microstruct.}\ }\textbf {\bibinfo {volume} {15}},\ \bibinfo
  {pages} {471--471} (\bibinfo {year} {1994})}\BibitemShut {NoStop}%
\bibitem [{\citenamefont {{Merle d\char39{}Aubign{\'e}}}\ \emph
  {et~al.}(1996)\citenamefont {{Merle d\char39{}Aubign{\'e}}}, \citenamefont
  {Wasiela}, \citenamefont {Mariette},\ and\ \citenamefont
  {Dietl}}]{Merle1996}%
  \BibitemOpen
  \bibfield  {author} {\bibinfo {author} {\bibfnamefont {Y.}~\bibnamefont
  {{Merle d\char39{}Aubign{\'e}}}}, \bibinfo {author} {\bibfnamefont
  {A.}~\bibnamefont {Wasiela}}, \bibinfo {author} {\bibfnamefont
  {H.}~\bibnamefont {Mariette}}, \ and\ \bibinfo {author} {\bibfnamefont
  {T.}~\bibnamefont {Dietl}},\ }\bibfield  {title} {\enquote {\bibinfo {title}
  {{P}olariton effects in multiple-quantum-well structures of
  $\mbox{CdTe}$/$\mbox{Cd}_{1-x}\mbox{Zn}_x\mbox{Te}$},}\ }\href {\doibase
  10.1103/PhysRevB.54.14003} {\bibfield  {journal} {\bibinfo  {journal} {Phys.
  Rev. B}\ }\textbf {\bibinfo {volume} {54}},\ \bibinfo {pages} {14003--14011}
  (\bibinfo {year} {1996})}\BibitemShut {NoStop}%
\bibitem [{\citenamefont {Prineas}\ \emph {et~al.}(2000)\citenamefont
  {Prineas}, \citenamefont {Ell}, \citenamefont {Lee}, \citenamefont
  {Khitrova}, \citenamefont {Gibbs},\ and\ \citenamefont {Koch}}]{Prineas2000}%
  \BibitemOpen
  \bibfield  {author} {\bibinfo {author} {\bibfnamefont {J.~P.}\ \bibnamefont
  {Prineas}}, \bibinfo {author} {\bibfnamefont {C.}~\bibnamefont {Ell}},
  \bibinfo {author} {\bibfnamefont {E.~S.}\ \bibnamefont {Lee}}, \bibinfo
  {author} {\bibfnamefont {G.}~\bibnamefont {Khitrova}}, \bibinfo {author}
  {\bibfnamefont {H.~M.}\ \bibnamefont {Gibbs}}, \ and\ \bibinfo {author}
  {\bibfnamefont {S.~W.}\ \bibnamefont {Koch}},\ }\bibfield  {title} {\enquote
  {\bibinfo {title} {{E}xciton-polariton eigenmodes in light-coupled
  $\mbox{In}_{0.04}\mbox{Ga}_{0.96}\mbox{As}/\mbox{GaAs}$ semiconductor
  multiple-quantum-well periodic structures},}\ }\href {\doibase
  10.1103/PhysRevB.61.13863} {\bibfield  {journal} {\bibinfo  {journal} {Phys.
  Rev. B}\ }\textbf {\bibinfo {volume} {61}},\ \bibinfo {pages} {13863--13872}
  (\bibinfo {year} {2000})}\BibitemShut {NoStop}%
\bibitem [{\citenamefont {{Goldberg}}\ \emph {et~al.}(2009)\citenamefont
  {{Goldberg}}, \citenamefont {{Deych}}, \citenamefont {{Lisyansky}},
  \citenamefont {{Shi}}, \citenamefont {{Menon}}, \citenamefont {{Tokranov}},
  \citenamefont {{Yakimov}},\ and\ \citenamefont
  {{Oktyabrsky}}}]{Goldberg2009}%
  \BibitemOpen
  \bibfield  {author} {\bibinfo {author} {\bibfnamefont {D.}~\bibnamefont
  {{Goldberg}}}, \bibinfo {author} {\bibfnamefont {L.~I.}\ \bibnamefont
  {{Deych}}}, \bibinfo {author} {\bibfnamefont {A.~A.}\ \bibnamefont
  {{Lisyansky}}}, \bibinfo {author} {\bibfnamefont {Z.}~\bibnamefont {{Shi}}},
  \bibinfo {author} {\bibfnamefont {V.~M.}\ \bibnamefont {{Menon}}}, \bibinfo
  {author} {\bibfnamefont {V.}~\bibnamefont {{Tokranov}}}, \bibinfo {author}
  {\bibfnamefont {M.}~\bibnamefont {{Yakimov}}}, \ and\ \bibinfo {author}
  {\bibfnamefont {S.}~\bibnamefont {{Oktyabrsky}}},\ }\bibfield  {title}
  {\enquote {\bibinfo {title} {{E}xciton-lattice polaritons in
  multiple-quantum-well-based photonic crystals},}\ }\href {\doibase
  10.1038/nphoton.2009.190} {\bibfield  {journal} {\bibinfo  {journal} {Nat.
  Photonics}\ }\textbf {\bibinfo {volume} {3}},\ \bibinfo {pages} {662--666}
  (\bibinfo {year} {2009})}\BibitemShut {NoStop}%
\bibitem [{\citenamefont {Askitopoulos}\ \emph {et~al.}(2011)\citenamefont
  {Askitopoulos}, \citenamefont {Mouchliadis}, \citenamefont {Iorsh},
  \citenamefont {Christmann}, \citenamefont {Baumberg}, \citenamefont
  {Kaliteevski}, \citenamefont {Hatzopoulos},\ and\ \citenamefont
  {Savvidis}}]{iorsh2011}%
  \BibitemOpen
  \bibfield  {author} {\bibinfo {author} {\bibfnamefont {A.}~\bibnamefont
  {Askitopoulos}}, \bibinfo {author} {\bibfnamefont {L.}~\bibnamefont
  {Mouchliadis}}, \bibinfo {author} {\bibfnamefont {I.}~\bibnamefont {Iorsh}},
  \bibinfo {author} {\bibfnamefont {G.}~\bibnamefont {Christmann}}, \bibinfo
  {author} {\bibfnamefont {J.~J.}\ \bibnamefont {Baumberg}}, \bibinfo {author}
  {\bibfnamefont {M.~A.}\ \bibnamefont {Kaliteevski}}, \bibinfo {author}
  {\bibfnamefont {Z.}~\bibnamefont {Hatzopoulos}}, \ and\ \bibinfo {author}
  {\bibfnamefont {P.~G.}\ \bibnamefont {Savvidis}},\ }\bibfield  {title}
  {\enquote {\bibinfo {title} {{B}ragg {P}olaritons: {S}trong {C}oupling and
  {A}mplification in an {U}nfolded {M}icrocavity},}\ }\href {\doibase
  10.1103/PhysRevLett.106.076401} {\bibfield  {journal} {\bibinfo  {journal}
  {Phys. Rev. Lett.}\ }\textbf {\bibinfo {volume} {106}},\ \bibinfo {pages}
  {076401} (\bibinfo {year} {2011})}\BibitemShut {NoStop}%
\bibitem [{\citenamefont {Chaldyshev}\ \emph
  {et~al.}(2011{\natexlab{b}})\citenamefont {Chaldyshev}, \citenamefont
  {Bolshakov}, \citenamefont {Zavarin}, \citenamefont {Sakharov}, \citenamefont
  {Lundin}, \citenamefont {Tsatsulnikov}, \citenamefont {Yagovkina},
  \citenamefont {Kim},\ and\ \citenamefont {Park}}]{chaldyshev2011b}%
  \BibitemOpen
  \bibfield  {author} {\bibinfo {author} {\bibfnamefont {V.~V.}\ \bibnamefont
  {Chaldyshev}}, \bibinfo {author} {\bibfnamefont {A.~S.}\ \bibnamefont
  {Bolshakov}}, \bibinfo {author} {\bibfnamefont {E.~E.}\ \bibnamefont
  {Zavarin}}, \bibinfo {author} {\bibfnamefont {A.~V.}\ \bibnamefont
  {Sakharov}}, \bibinfo {author} {\bibfnamefont {W.~V.}\ \bibnamefont
  {Lundin}}, \bibinfo {author} {\bibfnamefont {A.~F.}\ \bibnamefont
  {Tsatsulnikov}}, \bibinfo {author} {\bibfnamefont {M.~A.}\ \bibnamefont
  {Yagovkina}}, \bibinfo {author} {\bibfnamefont {Taek}\ \bibnamefont {Kim}}, \
  and\ \bibinfo {author} {\bibfnamefont {Youngsoo}\ \bibnamefont {Park}},\
  }\bibfield  {title} {\enquote {\bibinfo {title} {{O}ptical lattices of
  {I}n{G}a{N} quantum well excitons},}\ }\href {\doibase 10.1063/1.3670499}
  {\bibfield  {journal} {\bibinfo  {journal} {Appl. Phys. Lett.}\ }\textbf
  {\bibinfo {volume} {99}},\ \bibinfo {eid} {251103} (\bibinfo {year}
  {2011}{\natexlab{b}})}\BibitemShut {NoStop}%
\bibitem [{\citenamefont {Ammerlahn}\ \emph {et~al.}(2000)\citenamefont
  {Ammerlahn}, \citenamefont {Kuhl}, \citenamefont {Grote}, \citenamefont
  {Koch}, \citenamefont {Khitrova},\ and\ \citenamefont
  {Gibbs}}]{ammerlahn2000}%
  \BibitemOpen
  \bibfield  {author} {\bibinfo {author} {\bibfnamefont {D.}~\bibnamefont
  {Ammerlahn}}, \bibinfo {author} {\bibfnamefont {J.}~\bibnamefont {Kuhl}},
  \bibinfo {author} {\bibfnamefont {B.}~\bibnamefont {Grote}}, \bibinfo
  {author} {\bibfnamefont {S.~W.}\ \bibnamefont {Koch}}, \bibinfo {author}
  {\bibfnamefont {G.}~\bibnamefont {Khitrova}}, \ and\ \bibinfo {author}
  {\bibfnamefont {H.}~\bibnamefont {Gibbs}},\ }\bibfield  {title} {\enquote
  {\bibinfo {title} {{C}ollective radiative decay of light- and heavy-hole
  exciton polaritons in multiple-quantum-well structures},}\ }\href {\doibase
  10.1103/PhysRevB.62.7350} {\bibfield  {journal} {\bibinfo  {journal} {Phys.
  Rev. B}\ }\textbf {\bibinfo {volume} {62}},\ \bibinfo {pages} {7350--7356}
  (\bibinfo {year} {2000})}\BibitemShut {NoStop}%
\bibitem [{\citenamefont {{Volz}}\ \emph {et~al.}(2012)\citenamefont {{Volz}},
  \citenamefont {{Reinhard}}, \citenamefont {{Winger}}, \citenamefont
  {{Badolato}}, \citenamefont {{Hennessy}}, \citenamefont {{Hu}},\ and\
  \citenamefont {{Imamoglu}}}]{Volz2012}%
  \BibitemOpen
  \bibfield  {author} {\bibinfo {author} {\bibfnamefont {T.}~\bibnamefont
  {{Volz}}}, \bibinfo {author} {\bibfnamefont {A.}~\bibnamefont {{Reinhard}}},
  \bibinfo {author} {\bibfnamefont {M.}~\bibnamefont {{Winger}}}, \bibinfo
  {author} {\bibfnamefont {A.}~\bibnamefont {{Badolato}}}, \bibinfo {author}
  {\bibfnamefont {K.~J.}\ \bibnamefont {{Hennessy}}}, \bibinfo {author}
  {\bibfnamefont {E.~L.}\ \bibnamefont {{Hu}}}, \ and\ \bibinfo {author}
  {\bibfnamefont {A.}~\bibnamefont {{Imamoglu}}},\ }\bibfield  {title}
  {\enquote {\bibinfo {title} {{Ultrafast all-optical switching by single
  photons}},}\ }\href {\doibase vsdvsavs10.1038/nphoton.2012.181} {\bibfield
  {journal} {\bibinfo  {journal} {Nat. Photonics}\ } (\bibinfo {year} {2012}),\
  vsdvsavs10.1038/nphoton.2012.181}\BibitemShut {NoStop}%
\bibitem [{\citenamefont {Andreani}\ \emph {et~al.}(1998)\citenamefont
  {Andreani}, \citenamefont {Panzarini}, \citenamefont {Kavokin},\ and\
  \citenamefont {Vladimirova}}]{Andreani1998}%
  \BibitemOpen
  \bibfield  {author} {\bibinfo {author} {\bibfnamefont {Lucio~Claudio}\
  \bibnamefont {Andreani}}, \bibinfo {author} {\bibfnamefont {Giovanna}\
  \bibnamefont {Panzarini}}, \bibinfo {author} {\bibfnamefont {Alexey~V.}\
  \bibnamefont {Kavokin}}, \ and\ \bibinfo {author} {\bibfnamefont {Maria~R.}\
  \bibnamefont {Vladimirova}},\ }\bibfield  {title} {\enquote {\bibinfo {title}
  {{E}ffect of inhomogeneous broadening on optical properties of excitons in
  quantum wells},}\ }\href {\doibase 10.1103/PhysRevB.57.4670} {\bibfield
  {journal} {\bibinfo  {journal} {Phys. Rev. B}\ }\textbf {\bibinfo {volume}
  {57}},\ \bibinfo {pages} {4670--4680} (\bibinfo {year} {1998})}\BibitemShut
  {NoStop}%
\bibitem [{\citenamefont {Poddubny}\ \emph {et~al.}(2008)\citenamefont
  {Poddubny}, \citenamefont {Pilozzi}, \citenamefont {Voronov},\ and\
  \citenamefont {Ivchenko}}]{Poddubny2008prb}%
  \BibitemOpen
  \bibfield  {author} {\bibinfo {author} {\bibfnamefont {A.~N.}\ \bibnamefont
  {Poddubny}}, \bibinfo {author} {\bibfnamefont {L.}~\bibnamefont {Pilozzi}},
  \bibinfo {author} {\bibfnamefont {M.~M.}\ \bibnamefont {Voronov}}, \ and\
  \bibinfo {author} {\bibfnamefont {E.~L.}\ \bibnamefont {Ivchenko}},\
  }\bibfield  {title} {\enquote {\bibinfo {title} {{R}esonant
  $\mbox{F}$ibonacci quantum well structures in one dimension},}\ }\href
  {\doibase 10.1103/PhysRevB.77.113306} {\bibfield  {journal} {\bibinfo
  {journal} {Phys. Rev. B}\ }\textbf {\bibinfo {volume} {77}},\ \bibinfo
  {pages} {113306} (\bibinfo {year} {2008})}\BibitemShut {NoStop}%
\bibitem [{\citenamefont {Hendrickson}\ \emph {et~al.}(2008)\citenamefont
  {Hendrickson}, \citenamefont {Richards}, \citenamefont {Sweet}, \citenamefont
  {Khitrova}, \citenamefont {Poddubny}, \citenamefont {Ivchenko}, \citenamefont
  {Wegener},\ and\ \citenamefont {Gibbs}}]{Hendrickson2008}%
  \BibitemOpen
  \bibfield  {author} {\bibinfo {author} {\bibfnamefont {J.}~\bibnamefont
  {Hendrickson}}, \bibinfo {author} {\bibfnamefont {B.~C.}\ \bibnamefont
  {Richards}}, \bibinfo {author} {\bibfnamefont {J.}~\bibnamefont {Sweet}},
  \bibinfo {author} {\bibfnamefont {G.}~\bibnamefont {Khitrova}}, \bibinfo
  {author} {\bibfnamefont {A.~N.}\ \bibnamefont {Poddubny}}, \bibinfo {author}
  {\bibfnamefont {E.~L.}\ \bibnamefont {Ivchenko}}, \bibinfo {author}
  {\bibfnamefont {M.}~\bibnamefont {Wegener}}, \ and\ \bibinfo {author}
  {\bibfnamefont {H.~M.}\ \bibnamefont {Gibbs}},\ }\bibfield  {title} {\enquote
  {\bibinfo {title} {{E}xcitonic polaritons in $\mbox{F}$ibonacci
  quasicrystals},}\ }\href {\doibase 10.1364/OE.16.015382} {\bibfield
  {journal} {\bibinfo  {journal} {Opt. Express}\ }\textbf {\bibinfo {volume}
  {16}},\ \bibinfo {pages} {15382--15387} (\bibinfo {year} {2008})}\BibitemShut
  {NoStop}%
\bibitem [{\citenamefont {Dicke}(1954)}]{Dicke1954}%
  \BibitemOpen
  \bibfield  {author} {\bibinfo {author} {\bibfnamefont {R.~H.}\ \bibnamefont
  {Dicke}},\ }\bibfield  {title} {\enquote {\bibinfo {title} {{C}oherence in
  {S}pontaneous {R}adiation {P}rocesses},}\ }\href {\doibase
  10.1103/PhysRev.93.99} {\bibfield  {journal} {\bibinfo  {journal} {Phys.
  Rev.}\ }\textbf {\bibinfo {volume} {93}},\ \bibinfo {pages} {99} (\bibinfo
  {year} {1954})}\BibitemShut {NoStop}%
\bibitem [{\citenamefont {{Khitrova}}\ and\ \citenamefont
  {{Gibbs}}(2007)}]{Khitrova2007nat}%
  \BibitemOpen
  \bibfield  {author} {\bibinfo {author} {\bibfnamefont {G.}~\bibnamefont
  {{Khitrova}}}\ and\ \bibinfo {author} {\bibfnamefont {H.~M.}\ \bibnamefont
  {{Gibbs}}},\ }\bibfield  {title} {\enquote {\bibinfo {title} {{Q}uantum dots:
  {C}ollective radiance},}\ }\href {\doibase 10.1038/nphys532} {\bibfield
  {journal} {\bibinfo  {journal} {Nat. Physics}\ }\textbf {\bibinfo {volume}
  {3}},\ \bibinfo {pages} {84--86} (\bibinfo {year} {2007})}\BibitemShut
  {NoStop}%
\bibitem [{\citenamefont {Borrmann}(1950)}]{borrmann1950}%
  \BibitemOpen
  \bibfield  {author} {\bibinfo {author} {\bibfnamefont {G.}~\bibnamefont
  {Borrmann}},\ }\bibfield  {title} {\enquote {\bibinfo {title} {{D}ie
  {A}bsorption von {R}{\"o}ntgenstrahlen im {F}all der {I}nterferenz},}\ }\href
  {\doibase 10.1007/BF01329828} {\bibfield  {journal} {\bibinfo  {journal}
  {Zeitschrift f{\"u}r Physik}\ }\textbf {\bibinfo {volume} {127}},\ \bibinfo
  {pages} {297--323} (\bibinfo {year} {1950})}\BibitemShut {NoStop}%
\bibitem [{\citenamefont {{Voronov}}\ \emph {et~al.}(2007)\citenamefont
  {{Voronov}}, \citenamefont {{Ivchenko}}, \citenamefont {{Kosobukin}},\ and\
  \citenamefont {{Poddubny{\u i}}}}]{SpecFreqs2}%
  \BibitemOpen
  \bibfield  {author} {\bibinfo {author} {\bibfnamefont {M.~M.}\ \bibnamefont
  {{Voronov}}}, \bibinfo {author} {\bibfnamefont {E.~L.}\ \bibnamefont
  {{Ivchenko}}}, \bibinfo {author} {\bibfnamefont {V.~A.}\ \bibnamefont
  {{Kosobukin}}}, \ and\ \bibinfo {author} {\bibfnamefont {A.~N.}\ \bibnamefont
  {{Poddubny{\u i}}}},\ }\bibfield  {title} {\enquote {\bibinfo {title}
  {{S}pecific features in reflectance and absorbance spectra of one-dimensional
  resonant photonic crystals},}\ }\href {\doibase 10.1134/S1063783407090302}
  {\bibfield  {journal} {\bibinfo  {journal} {Fiz. Tverd. Tela}\ }\textbf
  {\bibinfo {volume} {49}},\ \bibinfo {pages} {1709} (\bibinfo {year}
  {2007})},\ \translation{Phys. Solid State {\bf 49}, 1792 (2007)}\BibitemShut
  {NoStop}%
\bibitem [{\citenamefont {Janot}(1994)}]{Janot}%
  \BibitemOpen
  \bibfield  {author} {\bibinfo {author} {\bibfnamefont {C.}~\bibnamefont
  {Janot}},\ }\href@noop {} {\emph {\bibinfo {title} {{Q}uasicrystals. {A}
  {P}rimer}}}\ (\bibinfo  {publisher} {Clarendon Press},\ \bibinfo {address}
  {Oxford, UK},\ \bibinfo {year} {1994})\BibitemShut {NoStop}%
\bibitem [{\citenamefont {Steurera}\ and\ \citenamefont
  {Haibacha}(2006)}]{cryst2006}%
  \BibitemOpen
  \bibfield  {author} {\bibinfo {author} {\bibfnamefont {W.}~\bibnamefont
  {Steurera}}\ and\ \bibinfo {author} {\bibfnamefont {T.}~\bibnamefont
  {Haibacha}},\ }\href@noop {} {\emph {\bibinfo {title} {{I}nternational
  {T}ables for {C}rystallography {V}olume {B}. {C}hapter 4.6.
  {R}eciprocal-space images of aperiodic crystals}}}\ (\bibinfo  {publisher}
  {International Union of Crystallography},\ \bibinfo {year}
  {2006})\BibitemShut {NoStop}%
\bibitem [{\citenamefont {Chumakov}\ \emph {et~al.}(1993)\citenamefont
  {Chumakov}, \citenamefont {Smirnov}, \citenamefont {Baron}, \citenamefont
  {Arthur}, \citenamefont {Brown}, \citenamefont {Ruby}, \citenamefont
  {Brown},\ and\ \citenamefont {Salashchenko}}]{chumakov1993}%
  \BibitemOpen
  \bibfield  {author} {\bibinfo {author} {\bibfnamefont {A.~I.}\ \bibnamefont
  {Chumakov}}, \bibinfo {author} {\bibfnamefont {G.~V.}\ \bibnamefont
  {Smirnov}}, \bibinfo {author} {\bibfnamefont {A.~Q.~R.}\ \bibnamefont
  {Baron}}, \bibinfo {author} {\bibfnamefont {J.}~\bibnamefont {Arthur}},
  \bibinfo {author} {\bibfnamefont {D.~E.}\ \bibnamefont {Brown}}, \bibinfo
  {author} {\bibfnamefont {S.~L.}\ \bibnamefont {Ruby}}, \bibinfo {author}
  {\bibfnamefont {G.~S.}\ \bibnamefont {Brown}}, \ and\ \bibinfo {author}
  {\bibfnamefont {N.~N.}\ \bibnamefont {Salashchenko}},\ }\bibfield  {title}
  {\enquote {\bibinfo {title} {{R}esonant diffraction of synchrotron radiation
  by a nuclear multilayer},}\ }\href {\doibase 10.1103/PhysRevLett.71.2489}
  {\bibfield  {journal} {\bibinfo  {journal} {Phys. Rev. Lett.}\ }\textbf
  {\bibinfo {volume} {71}},\ \bibinfo {pages} {2489--2492} (\bibinfo {year}
  {1993})}\BibitemShut {NoStop}%
\bibitem [{\citenamefont {Poddubny}\ \emph {et~al.}(2009)\citenamefont
  {Poddubny}, \citenamefont {Pilozzi}, \citenamefont {Voronov},\ and\
  \citenamefont {Ivchenko}}]{poddubny2009}%
  \BibitemOpen
  \bibfield  {author} {\bibinfo {author} {\bibfnamefont {A.~N.}\ \bibnamefont
  {Poddubny}}, \bibinfo {author} {\bibfnamefont {L.}~\bibnamefont {Pilozzi}},
  \bibinfo {author} {\bibfnamefont {M.~M.}\ \bibnamefont {Voronov}}, \ and\
  \bibinfo {author} {\bibfnamefont {E.~L.}\ \bibnamefont {Ivchenko}},\
  }\bibfield  {title} {\enquote {\bibinfo {title} {{E}xciton-polaritonic
  quasicrystalline and aperiodic structures},}\ }\href {\doibase
  10.1103/PhysRevB.80.115314} {\bibfield  {journal} {\bibinfo  {journal}
  {Phys.~Rev.~B}\ }\textbf {\bibinfo {volume} {80}},\ \bibinfo {eid} {115314}
  (\bibinfo {year} {2009})}\BibitemShut {NoStop}%
\bibitem [{\citenamefont {Chumakov}\ \emph {et~al.}(1999)\citenamefont
  {Chumakov}, \citenamefont {Niesen}, \citenamefont {Nagy},\ and\ \citenamefont
  {Alp}}]{chumakov1999}%
  \BibitemOpen
  \bibfield  {author} {\bibinfo {author} {\bibfnamefont {A.I.}\ \bibnamefont
  {Chumakov}}, \bibinfo {author} {\bibfnamefont {L.}~\bibnamefont {Niesen}},
  \bibinfo {author} {\bibfnamefont {D.L.}\ \bibnamefont {Nagy}}, \ and\
  \bibinfo {author} {\bibfnamefont {E.E.}\ \bibnamefont {Alp}},\ }\bibfield
  {title} {\enquote {\bibinfo {title} {{N}uclear resonant scattering of
  synchrotron radiation by multilayer structures},}\ }\href@noop {} {\bibfield
  {journal} {\bibinfo  {journal} {Hyperfine Interactions}\ }\textbf {\bibinfo
  {volume} {123/124}},\ \bibinfo {pages} {427--454} (\bibinfo {year}
  {1999})}\BibitemShut {NoStop}%
\bibitem [{\citenamefont {Vladimirova}\ \emph {et~al.}(1998)\citenamefont
  {Vladimirova}, \citenamefont {Ivchenko},\ and\ \citenamefont
  {Kavokin}}]{vladimirova1998}%
  \BibitemOpen
  \bibfield  {author} {\bibinfo {author} {\bibfnamefont {M.~R.}\ \bibnamefont
  {Vladimirova}}, \bibinfo {author} {\bibfnamefont {E.~L.}\ \bibnamefont
  {Ivchenko}}, \ and\ \bibinfo {author} {\bibfnamefont {A.~V.}\ \bibnamefont
  {Kavokin}},\ }\bibfield  {title} {\enquote {\bibinfo {title} {{E}xciton
  polaritons in long-period quantum-well structures},}\ }\href {\doibase
  10.1134/1.1187364} {\bibfield  {journal} {\bibinfo  {journal} {Fiz. Tekh.
  Poluprovodn.}\ }\textbf {\bibinfo {volume} {32}},\ \bibinfo {pages} {101}
  (\bibinfo {year} {1998})},\ \translation{Semiconductors {\bf 32}, 90
  (1998)}\BibitemShut {NoStop}%
\bibitem [{\citenamefont {Landau}\ and\ \citenamefont
  {Lifshit͡s}(1975)}]{LL2}%
  \BibitemOpen
  \bibfield  {author} {\bibinfo {author} {\bibfnamefont {L.D.}\ \bibnamefont
  {Landau}}\ and\ \bibinfo {author} {\bibfnamefont {E.M.}\ \bibnamefont
  {Lifshit͡s}},\ }\href@noop {} {\emph {\bibinfo {title} {{T}he {C}lassical
  {T}heory of {F}ields}}},\ Course of Theoretical Physics, v. 2\ (\bibinfo
  {publisher} {Butterworth-Heinemann},\ \bibinfo {year} {1975})\BibitemShut
  {NoStop}%
\end{thebibliography}
\begin{figure}[t!]
\includegraphics[width=0.49\textwidth]{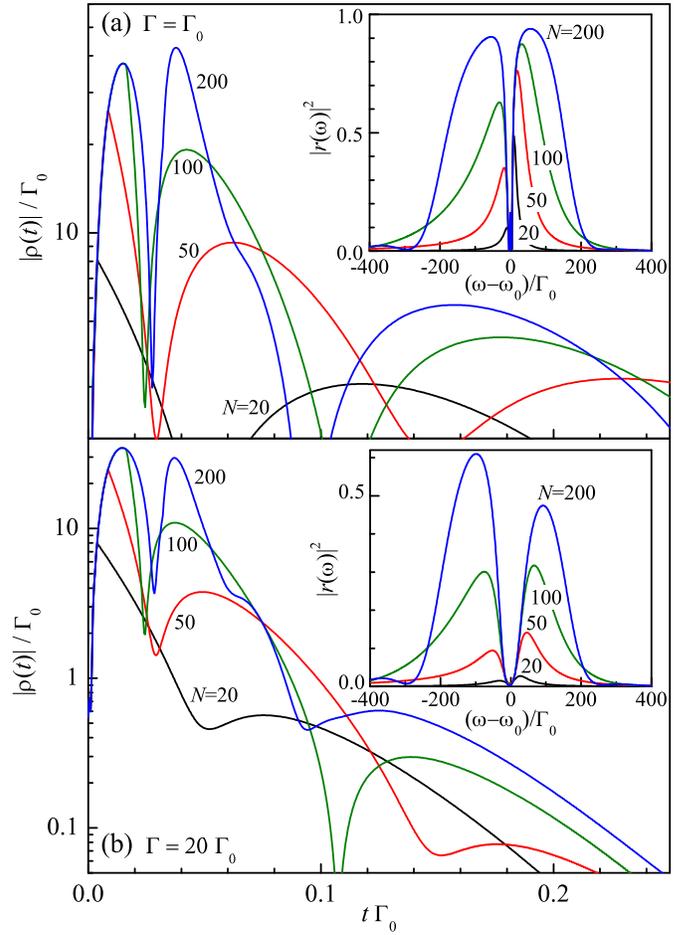}
\caption{(Color online) The envelopes of reflected pulses calculated for the gradient structures with various  numbers of QWs, $\zeta=5$, and $\omega_0/\Gamma_0 = 4 \times 10^4$. Figures~(a) and~(b) correspond to weak, $\Gamma=\Gamma_0$, and strong, $\Gamma=20 \Gamma_0$, non-radiative decay, respectively. Insets show the reflection spectra of the same QW structures in the frequency domain.}
\label{figure6}
\end{figure}

The envelopes of reflected pulses $\rho_N(t)$ in response to the ultrashort excitation pulse are presented in Figs.~\ref{figure6}(a) and~\ref{figure6}(b) for different gradient structures. The curves are calculated numerically by the transfer-matrix technique. Similarly to the optical response of the Fibonacci structures, the pulse envelope exhibits beatings even for structures with rather small QW numbers $N$. However, the pattern of the beatings is different from that in the Fibonacci structures,
and the pulse envelopes appear to be highly sensitive to the non-radiative decay rate $\Gamma$.
The reflection spectra of the same gradient structures in the frequency domain $r_N(\omega)$ are shown in insets in Figs.~\ref{figure6}(a) and~\ref{figure6}(b). The spectra are much more asymmetric around the exciton frequency $\omega_0$ as compared to the spectra of the resonance periodic or Fibonacci structures, see insets in Figs.~\ref{figure2} and~\ref{figure5}. The spectrum asymmetry is caused by the detuning of the structure front part from the Bragg condition. Note that the similar effect occurs in periodic structures detuned from the Bragg resonance,~\cite{vladimirova1998,Voronov2004} however, the integral reflectivity of the gradient structures can be larger than that of non-Bragg periodic ones because the main part of the gradient structures is tuned to the Bragg condition. Another feature of the gradient QW structures is the narrow dip in the reflection spectrum near the exciton resonance. Our calculations indicate that the dip is highly sensitive to the non-radiative decay and disappears at $\Gamma=0$.

\section{Polarization conversion}\label{sec:polarized}

The resonant photonic crystals and quasicrystals can change the polarization state of an optical pulse if the exciton levels forming the photon band structure are split. The splitting can be induced by an external magnetic or electric field, strain, or intrinsic in-plane anisotropy of the QWs, e.g., in low-symmetry structures.\cite{Ivchenko_book} In this section, we consider the polarization conversion caused by the magnetic field applied along the structure axis (Faraday geometry). The field splits the exciton level in each QW into two states interacting with the right- and left-handed circular polarized radiation. The corresponding resonance frequencies are given by $\omega_0^{(\pm)}=\omega_0 \pm \Omega_L /2$, where $\Omega_L$ is the Larmor frequency corresponding to the Zeeman splitting. The level splitting results in the difference of the reflection spectra for the right- and left-handed circularly polarized light,
which leads, in turn, to the change in the reflected pulse polarization with respect to the excitation pulse. For definiteness, we focus on the effect of the magnetic field on the resonance frequencies and neglect a possible change in the exciton oscillator strength.

To describe the reflected pulse polarization, we introduce the Stokes parameters $S_1$, $S_2$, and $S_3$, which give
the degree of linear polarization along the axes $x$ and $y$ perpendicular to the growth direction $z$, the degree of linear polarization along the axes rotated by $45^\circ$, and the degree of circular polarization, respectively,\cite{LL2} by
\begin{eqnarray}\label{Stokes}
S_1(t) &=& \langle |E_{R,x}(t)|^2 - |E_{R,y}(t)|^2 \rangle  \:, \nonumber \\
S_2(t) &=& \langle E_{R,x}(t) E_{R,y}^*(t) + E_{R,y}(t) E_{R,x}^*(t) \rangle  \:, \nonumber \\
S_3(t) &=& - {\rm i} \langle E_{R,x}(t) E_{R,y}^*(t) - E_{R,y}(t) E_{R,x}^*(t) \rangle \:. \;\;
\end{eqnarray}
Here, the angular brackets denote the averaging over the time domain longer than $2\pi/\omega_0$
but shorter than $1/\Delta$ and $1/\Omega_L$.
Note, that the sign of $S_3$ takes into account the fact that the reflected wave moves in the $-z$ direction. The intensity of the reflected wave is given by
\begin{equation}\label{Intensity}
I_R(t) = \langle |E_{R,x}(t)|^2 + |E_{R,y}(t)|^2 \rangle \:.
\end{equation}

Straightforward calculation shows that the Stokes parameters and intensity of the reflected wave have the form
\[
S_1(t) = {\rm Re} [\rho_{+} \rho_{-}^*] (J_{xx}-J_{yy}) +  {\rm Im} [\rho_{+}\rho_{-}^*] (J_{xy}+J_{yx}) \,,
\]
\[
S_2(t) = {\rm Re} [\rho_{+} \rho_{-}^*] (J_{xy}+J_{yx}) -  {\rm Im} [\rho_{+}\rho_{-}^*] (J_{xx}-J_{yy}) \,,
\]
\[
S_3(t) = \frac{|\rho_+|^2 \hspace{-1mm} + \hspace{-1mm}|\rho_-|^2}{2} {\rm i} (J_{yx}-J_{xy}) - \frac{|\rho_+|^2 \hspace{-1mm} - \hspace{-1mm}|\rho_-|^2}{2} (J_{xx}+J_{yy}) \,,
\]
\begin{equation}
I_R(t) = \frac{|\rho_+|^2 \hspace{-1mm} + \hspace{-1mm} |\rho_-|^2}{2} (J_{xx}+J_{yy}) - \frac{|\rho_+|^2 \hspace{-1mm} - \hspace{-1mm} |\rho_-|^2}{2} {\rm i }(J_{yx}-J_{xy}) \,,
\end{equation}
where $\rho_{+} = \rho_{N,+}(t)$ and $\rho_{-} = \rho_{N,-}(t)$ are the response functions for the right- and left-handed circularly polarized radiation, respectively, and $J_{\alpha\beta}$ ($\alpha,\beta=x,y$) is the tensor determined by the incident wave,
\begin{equation}
J_{\alpha\beta} = \int_{-\infty}^{+\infty} E_{0,\alpha}(t) {\rm e}^{{\rm i}\omega_0 t} dt \int_{-\infty}^{+\infty} E_{0,\beta}^*(t') {\rm e}^{-{\rm i}\omega_0 t'} dt' \:.
\end{equation}
In the particular case of the incident wave linearly polarized along the $x$ axis, the Stokes parameters and intensity of the reflected wave assume the form
\begin{equation}\label{Stokes_final}
S_1(t) = {\rm Re} [\rho_{-} \rho_{+}^*] J_{xx} \:, \;\; S_2(t) = {\rm Im} [\rho_{-} \rho_{+}^*] J_{xx} \:,
\end{equation}
\[
S_3(t) = \frac{|\rho_{-}|^2 - |\rho_{+}|^2}{2} J_{xx} \:, \;\; I_R(t) = \frac{|\rho_+|^2 + |\rho_-|^2}{2} J_{xx} \:.
\]

\begin{figure}[t!]
\includegraphics[width=0.45\textwidth]{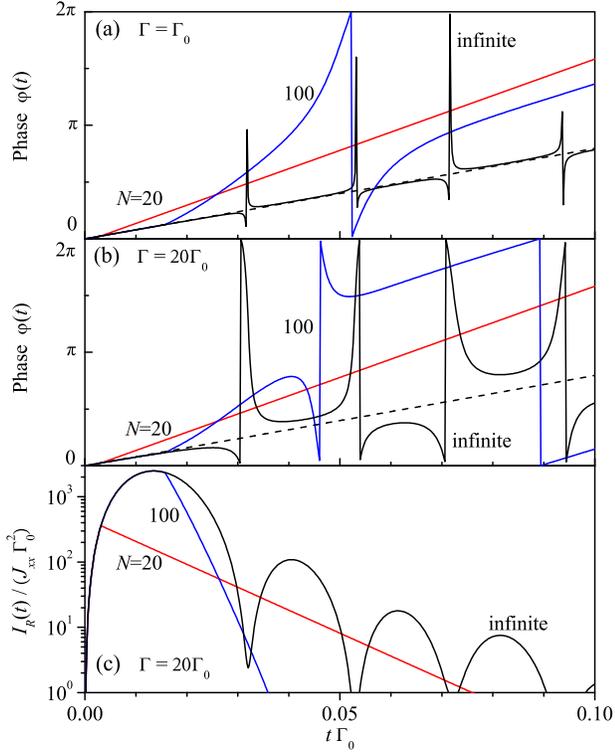}
\caption{(Color online) (a) and (b) Time dependences of the phase $\varphi$, which determines the reflected pulse polarization plane, for the periodic structures with various numbers of QWs, $\omega_0/\Gamma_0 = 4 \times 10^4$ and $\Omega_L=50 \Gamma_0$. Solid curves are the result of numerical calculation after Eqs.~(\ref{Stokes_final}), dashed curves show the linear dependence $\varphi(t)=\Omega_L t/2$. Figures~(a) and~(b) correspond to weak, $\Gamma=\Gamma_0$, and strong, $\Gamma=20 \Gamma_0$, non-radiative decay, respectively. (c) Time dependence of the reflected pulse intensities.}
\label{figure7}
\end{figure}
The excitation of a periodic QW structure by the linearly polarized pulse results in the linearly polarized response, whose polarization plane slowly rotates in time due to the Zeeman splitting of exciton states.
The circular polarization does not occur because, for periodic structures, $|\rho_{+}(t)|=|\rho_{-}(t)|$ which follows from the relation
\begin{equation}
r_{+}(\omega)=r_{-}^*(2\omega_0-\omega) \;.
\end{equation}
The latter can be readily seen from the transfer matrices
through the QW structures detuned from the Bragg condition by the magnetic field. Thus, the polarization state of the reflected signal can be completely described by the slowly varying phase $\varphi(t)$,
\begin{equation}\label{spectrum_symmetry}
S_1(t) + {\rm i} S_2(t) = I_R(t) \exp[{\rm i} \varphi(t)] \:,
\end{equation}
which determines the light polarization plane. Figures~\ref{figure7}(a) and~\ref{figure7}(b) show the dependences $\varphi(t)$ calculated numerically for structures with various QW numbers, weak [Fig.~\ref{figure7}(a)] and strong [Fig.~\ref{figure7}(b)] non-radiative decays. 
For comparison, Fig.~\ref{figure7}(c) presents the time dependences of intensities $I_R(t)$ of the same reflected pulses. One can see that, in contrast to naive expectation that the polarization plane rotates with the Larmor frequency, the dependence $\varphi(t)$ is different for structures with different QW numbers $N$ and can be non-linear. The simple law $\varphi(t) = \Omega_L t$ is valid only for structures with small QW numbers, $N \ll \sqrt{\omega_0/\Gamma_0},\omega_0/\Omega_L$, where the reflection coefficients $r_{\pm}(\omega)$ are given by Lorentzian Eq.~(\ref{eq:rsr}) with $\omega_0$ being replaced by $\omega_0 \pm \Omega_L/2$, respectively.
The reflection coefficients $r_{\pm}(\omega)$ of the semi-infinite structures in the magnetic field are described by Eq.~(\ref{eq:rinf}), where $w=(\omega-\omega_0)/\Delta$ and $\tilde{w}=(\omega-\omega_0\mp\Omega_L/2 + {\rm i}\Gamma)/\Delta$. At $\Gamma=0$, they fulfill the equation $r_+(\omega + \Omega_L/4 )=r_-(\omega - \Omega_L/4)$. Consequently, the frequency of the plane rotation becomes twice smaller and given by $\Omega_L/2$. The linear function $\varphi(t) = \Omega_L t/2$ shown by dashed curve describes well the numerically calculated dependence $\varphi(t)$ for the semi-infinite QW structure with weak non-radiative decay, see Fig.~\ref{figure7}(a). The points where the phase abruptly changes are close to the points of minima in the pulse intensity. The reduction of the plane rotation frequency in the semi-infinite structure by the factor of two as compared to that in short structures can be interpreted by the formation of polariton (coupled exciton and photon) modes in the semi-infinite structure. The magnetic field affects the exciton part of the polariton only, which leads to the Zeeman splitting of polaritons that is twice as small as the Zeeman splitting of free excitons.

\begin{figure}[t!]
\includegraphics[width=0.45\textwidth]{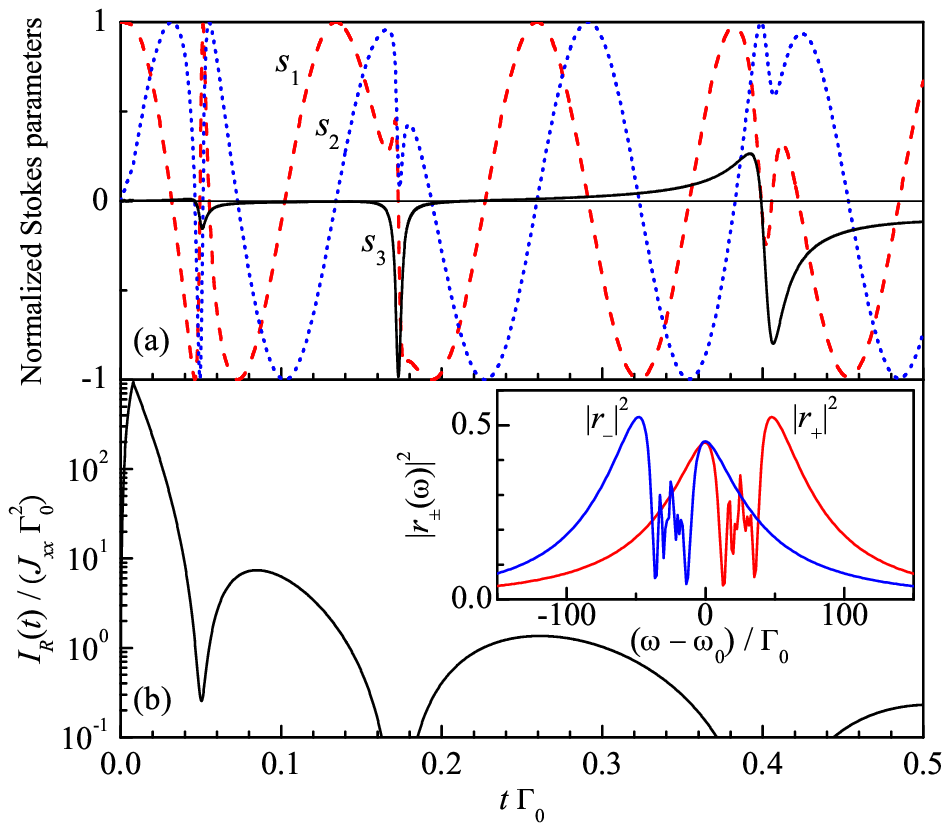}
\caption{(Color online) (a) Time dependence of the normalized Stokes parameters determining the reflected pulse polarization for the Fibonacci structure with $N=50$ QWs, $\omega_0/\Gamma_0 = 4 \times 10^4$, $\Gamma=\Gamma_0$, and $\Omega_L=50 \Gamma_0$. (b) Time dependence of the reflected pulse intensity. The inset shows the reflection spectra for the left-handed and right-handed circularly polarized light in the frequency domain.}
\label{figure8}
\vspace{0.5cm}
\includegraphics[width=0.45\textwidth]{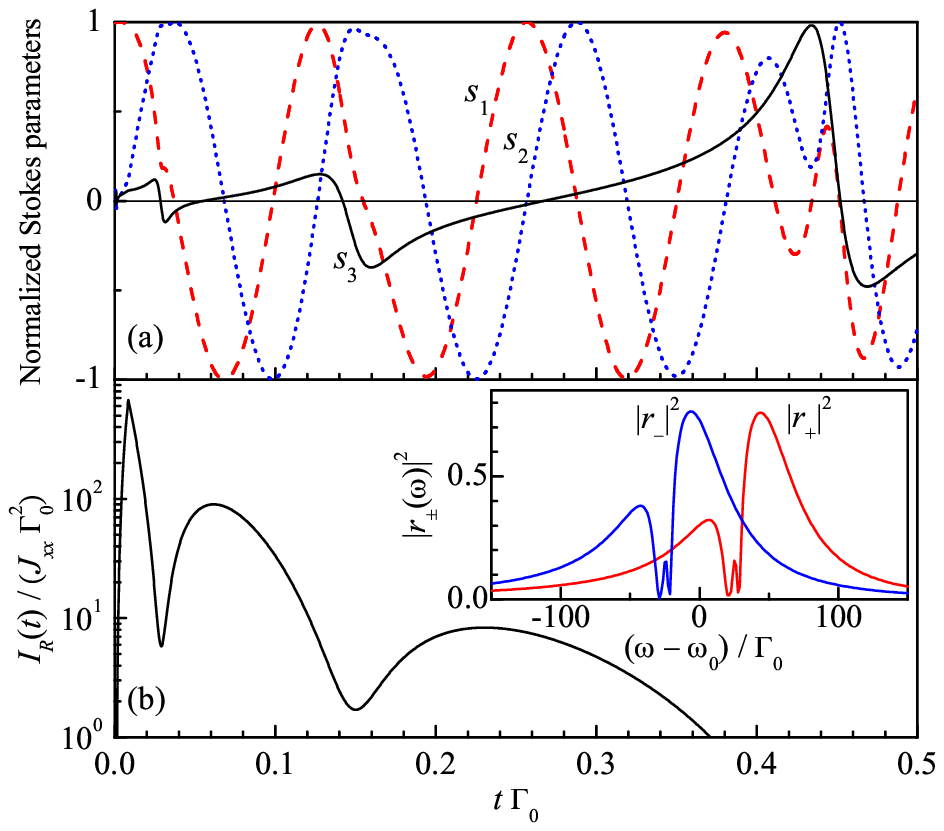}
\caption{(Color online) (a) Time dependence of the normalized Stokes parameters determining the reflected pulse polarization for the gradient structure with $N=50$ QWs, $\zeta=5$, $\omega_0/\Gamma_0 = 4 \times 10^4$, $\Gamma=\Gamma_0$, and $\Omega_L = 50 \Gamma_0$. (c) Time dependence of the reflected pulse intensity. The inset shows the reflection spectra for the left-handed and right-handed circularly polarized light in the frequency domain.}
\label{figure9}
\end{figure}
Figures~\ref{figure8} and~\ref{figure9} show the time dependence of the normalized Stokes parameters
$s_j(t)=S_j(t)/I_R(t)$ ($j=1,2,3$) and intensity $I_R(t)$ of the optical pulse reflected from the Fibonacci and gradient QW structures, respectively, in the response to excitation with the short pulse linearly polarized along the $x$ axis. In aperiodic systems, the relation~(\ref{spectrum_symmetry}), valid for periodic structures, does not hold anymore. Therefore, the conversion of linear to circular polarization becomes possible, i.e., $s_3(t) \neq 0$, see Figs.~\ref{figure8}(a) and~\ref{figure9}(a). However, the efficiency of such a process drastically depends on the QW spatial arrangement. In particular, the conversion efficiency is rather low in the Fibonacci structures. This is caused by the fact that Eq.~(\ref{spectrum_symmetry}) still holds for the smooth part of the reflection spectra, see inset in Fig.~\ref{figure8}(b). Thus, the linear-to-circular polarization conversion occurs due to the fine structure of the reflection spectrum central part and can be theoretically described only beyond the two-wave approximation. In the gradient structures, the reflection spectra are strongly asymmetric [Fig.~\ref{figure9}(b)], which leads to a more efficient polarization conversion.

\section{Summary}\label{sec:summary}

We have developed the microscopic theory of the optical response of one-dimensional photonic structures formed by periodic, Fibonacci, or gradient sequences of quantum wells to ultrashort polarized optical pulses. It is shown that the response depends on the structure optical length as well as the QW spatial arrangement. 
The excitation of short periodic QW structures tuned to the Bragg condition with an ultrashort pulse leads to formation of the reflected signal with a sharp front followed by a slow exponential decay. Both the signal amplitude and the decay rate increase with the QW number. 
In contrast, the reflected pulse envelope in long and, particular, semi-infinite structures exhibits Bessel oscillations with the period determined by the stop band width. Moreover, the trailing edge of the pulse reflected from the semi-infinite structure has a power asymptotics, rather than exponential one, even in the presence of non-radiative exciton decay in QWs. The qualitative change in the optical response of short and long structures is caused by the transition from superradiant to photonic-crystalline regime. In the Fibonacci or gradient QW structures, the reflected pulses exhibit oscillations even for structures with rather small QW numbers operating in the superradiant regime.

The duration of reflected pulses is shown to non-monotonically depend on the QW number, both for periodic and aperiodic structures. The duration decreases with the QW number in short structures, reaches the minimum in the area of the superradiant-photonic-crystalline regime transition, and then again increases in long structures. This behavior suggests the optimal number of quantum wells for designing the fast-reflecting Bragg structures.

We also have studied the conversion of light polarization in the external magnetic field applied along the growth direction. In this geometry, the excitation of a periodic QW structure by the linearly polarized pulse results in the linearly polarized response, whose polarization plane slowly varies in time (time-resolved magneto-optical Kerr effect). The plane rotation frequency in short structures is given by the Larmor frequency of excitons while in long structures the rotation frequency becomes twice smaller, which is related to the formation of polariton modes. In the Fibonacci and gradient structures, the reflection of linearly polarized pulse leads to both the rotation of the polarization plane and the appearance of partial circular polarization. The linear-to-circular polarization conversion drastically depends on the QW spatial arrangement vanishing for the periodic Bragg structures and is more pronounced in the gradient QW structures.

\paragraph*{Acknowledgments.} The authors acknowledge fruitful discussions with
E.L. Ivchenko. This work was supported by the RFBR, 
RF President Grants MD-2062.2012.2 and NSh-5442.2012.2, EU projects ``Spinoptronics'' and
``POLAPHEN'', and the Foundation ``Dynasty''.

\end{document}